\documentclass[11pt]{article}
\usepackage{amssymb,amsmath,amsfonts}
\usepackage{graphicx}
\usepackage{graphics}
\usepackage{eepic,epsfig}

\@addtoreset{equation}{section}

\makeatother

\begin{document}

\title{Quantum vacuum effects in braneworlds on AdS bulk}
\author{ Aram A. Saharian\thanks{%
E-mail: saharian@ysu.am } \vspace{0.3cm} \\
\textit{Department of Physics, Yerevan State University,}\\
\textit{1 Alex Manoogian Street, 0025 Yerevan, Armenia }}
\maketitle

\begin{abstract}
We review the results of investigations for brane-induced effects on the
local properties of quantum vacuum in background of AdS spacetime. Two
geometries are considered: a brane parallel to the AdS boundary and a brane
intersecting the AdS boundary. For both these cases the contribution in the
vacuum expectation value (VEV) of the energy-momentum tensor is separated
explicitly and its behavior in various asymptotic regions of the parameters
is studied. It is shown that the influence of the gravitational field on the
local properties of the quantum vacuum is essential at distance from the
brane larger than the AdS curvature radius. In the geometry with a brane
parallel to the AdS boundary the VEV of the energy-momentum tensor is
considered for scalar field with the Robin boundary condition, for Dirac
field with the bag boundary condition and for the electromagnetic field. In
the latter case two types of boundary conditions are discussed. The first
one is a generalization of the perfect conductor boundary condition and the
second one corresponds to the confining boundary condition used in QCD for
gluons. For the geometry of a brane intersecting the AdS boundary, the case
of a scalar field is considered. The corresponding energy-momentum tensor,
apart from the diagonal components, has nonzero off-diagonal component. As a
consequence of the latter, in addition to the normal component, the Casimir
force acquires a component parallel to the brane.
\end{abstract}

\bigskip

\section{Introduction}

\label{sec:introd}

Quantum field theory in anti-de Sitter (AdS) spacetime is an active area of
research. This activity is motivated by several reasons. First of all, the
corresponding geometry is maximally symmetric and a sufficiently large
number of problems are exactly solvable on its background. This helps to
give an idea of the influence of the classical gravitational field on
quantum phenomena in more complicated geometries. Qualitatively new features
in the dynamics of quantum fields on the AdS background are related to the
lack of global hyperbolicity and to the existence of both regular and
irregular modes. In particular, boundary conditions on propagating fields
need to be imposed at timelike infinity to prevent loss of unitarity. The
different boundary conditions define different theories in the bulk. Another
new feature, that has no analogue in Minkowskian field theories, is related
to the possibility of regularization for infrared divergences in interacting
field theories without reducing the symmetries \cite{Call90}. This is
closely related to the natural length scale of the AdS spacetime. The high
interest to the AdS geometry is also related to its natural appearance as a
ground state in supergravity and as the near horizon geometry for extremal
black holes, black strings and domain walls.

The AdS spacetime plays a fundamental role in two exciting developments of
contemporary theoretical physics. The first one, the AdS/CFT correspondence
(for reviews see \cite{Ahar00,Nast15,Ammo15}), establishes duality between
string theories or supergravity in the AdS bulk and a conformal field theory
localized on the AdS boundary. This duality provides an interesting
possibility for the investigation of nonperturbative effects in both sides
of the correspondence by using the weak coupling regime of the dual theory.
The recent developments include applications in various condensed matter
systems such as holographic superconductors and topological insulators \cite%
{Pire14,Zaan15}. The second development is related to various types of
braneworld models with large extra dimensions \cite{Maar10}. In the
corresponding setup the standard model fields are localized on a brane
embedded in a higher dimensional AdS spacetime. Braneworlds naturally appear
in string/M-theory context and have been initially proposed for a
geometrical resolution of the hierarchy problem between the electroweak and
Planck energy scales. They provide an alternative framework to address the
problems in particle physics and cosmology from different perspectives.

An inherent feature of field theories in AdS/CFT correspondence and in
braneworld models is the need to impose boundary conditions on fields
propagating in the AdS bulk. They include the conditions on the AdS boundary
and the conditions on the branes in braneworld scenario. In particular, in
braneworld models of the Randall-Sundrum type the boundary conditions on the
branes are dictated by the $Z_{2}$-symmetry. In quantum field theory, the
boundary conditions modify the spectrum of the zero-point fluctuations and,
as a consequence, the vacuum expectation values (VEVs) are shifted by an
amount that depends on the bulk and boundary geometries and also on the
boundary conditions. This is the well-known Casimir effect \cite{Most97}-%
\cite{Casi11}. In braneworld models the Casimir forces acting on the branes
may provide a mechanism for stabilization of the brane location (for
mechanisms of moduli stabilization in warped geometries see, e.g., \cite%
{Gold99}-\cite{Fuji20} and references therein). In particular, this
stabilization is required to prevent the variations of physical constants on
the branes. In addition, the quantum effects of bulk fields generate a
cosmological constant on the brane. Motivated by these points, the Casimir
effect in braneworld models on the AdS bulk, with branes parallel to the AdS
boundary, has been investigated for scalar \cite{Fabi00}-\cite{Haba19},
fermionic \cite{Flac01b}-\cite{Eliz13} and vector fields \cite{Garr03}-\cite%
{Saha20}. The models with de Sitter branes have been discussed in \cite%
{Noji00d}-\cite{Pujo05}. The brane-induced quantum vacuum effects in AdS
spacetime with additional compact subspaces were considered in \cite{Flac03a}%
-\cite{Fran08}.

The main part of the papers on the Casimir effect in the AdS bulk consider
global quantities, such as the Casimir energy and the forces acting on the
branes. The local quantities carry more detailed information about the
properties of the quantum vacuum. In particular, the expectation value\ of
the energy-momentum tensor is of special interest. It appears as the source
in the semiclassical Einstein equations and therefore plays an important
role in modelling self-consistent dynamics involving the gravitational
field. The VEV of the energy-momentum tensor for a conformally coupled
scalar field in conformally-flat geometries has been investigated in \cite%
{Saha03}. Massive scalar fields with general curvature coupling in the
geometry of two branes on AdS bulk were considered in \cite{Knap04,Saha05}.
The Casimir densities for a $Z_{2}$-symmetric thick brane for the general
case of static plane symmetric interior structure have been discussed in
\cite{Saha07}. The VEVs of the energy-momentum tensor for Dirac spinor field
and for the electromagnetic field are investigated in \cite{Shao10,Eliz13}
and \cite{Saha16}-\cite{Saha20}. The geometry of a brane intersecting the
AdS boundary has been considered in \cite{Beze15}. For a scalar field with
general curvature coupling, the background AdS geometry with an additional
compact subspace is discussed in \cite{Saha06a,Saha06b}.

In the present paper we review the results for the VEV of the
energy-momentum tensor in the geometry of a single brane on background of $%
(D+1)$-dimensional AdS spacetime. The cases of scalar, Dirac and
electromagnetic fields will be considered. The organization of the paper is
as follows. In section \ref{sec:Modes}, for a planar brane parallel to the
AdS boundary, we consider complete sets of orthonormalized mode functions
for both the regions between the AdS boundary and the brane and between the
brane and the AdS horizon. The VEVs of the energy-momentum tensors for
scalar, Dirac and electromagnetic fields are investigated in section \ref%
{sec:EMT}. The behavior in the asymptotic regions of the parameters is
discussed in detail. Section \ref{sec:Surf} considers the VEV of the surface
energy-momentum tensor for a scalar field on a brane parallel to the AdS
boundary. Section \ref{sec:PerpBr} is devoted to the study of the vacuum
energy-momentum tensor for a scalar field in the geometry with a brane
perpendicular to the AdS boundary. The main results are summarized in
section \ref{sec:Conc}.

\section{Mode functions in the geometry with a brane parallel to the AdS
boundary}

\label{sec:Modes}

In Poincar\'{e} coordinates $(x^{0}=t,x^{1},\ldots ,x^{D-1},y)$, with $%
-\infty <x^{i},y<+\infty $, $i=0,1,\ldots ,D-1$, the line element for the $%
(D+1)$-dimensional AdS spacetime is presented as
\begin{equation}
ds^{2}=e^{-2y/a}\left[ \left( dx^{0}\right) ^{2}-\left( dx^{1}\right)
^{2}-\cdots -\left( dx^{D-1}\right) ^{2}\right] -dy^{2},  \label{metric}
\end{equation}%
where $a$ is the curvature radius of the background geometry sourced by a
negative cosmological constant $\Lambda =-D(D-1)/(2a^{2})$. For the
curvature scalar and the Ricci tensor one has $R=-D(D+1)/a^{2}$ and $R_{\mu
\rho }=-Dg_{\mu \rho }/a^{2}$, with the metric tensor $g_{\mu \rho }$
defined from (\ref{metric}). Introducing a new coordinate $z$, $0\leqslant
z<\infty $, in accordance with $z=ae^{y/a}$, the line element is written in
conformally flat form%
\begin{equation}
ds^{2}=g_{\mu \rho }dx^{\mu }dx^{\rho }=(a/z)^{2}\eta _{\mu \rho }dx^{\mu
}dx^{\rho },  \label{metric2}
\end{equation}%
where $x^{D}=z$ and $\eta _{\mu \rho }=\mathrm{diag}(1,-1,\ldots ,-1)$ is
the metric tensor for $(D+1)$-dimensional Minkowski spacetime. The
hypersurfaces $z=0$ and $z=\infty $ correspond to the AdS boundary and
horizon, respectively. In what follows we will work in the coordinate system
defined by (\ref{metric2}).

We are interested in the effects on the local properties of the quantum
vacuum induced by a codimension one brane. First we consider the case where
the brane is parallel to the AdS boundary and is located at $z=z_{0}$. It
divides the space into two subspaces: the region between the AdS boundary
and the brane, $0\leq z\leq z_{0}$ (L-region) and the region between the
brane and AdS horizon, $z_{0}\leq z<\infty $ (R-region). The brane has a
nonzero extrinsic curvature tensor and, as a consequence, the properties of
the vacuum state in the L- and R-regions are different. The evaluation of
the VEVs for local physical observables in those regions requires different
procedures and we will discuss them separately. The VEVs are presented in
the form of mode-sums over complete set of mode functions for quantum fields
and we start by considering the modes for scalar, Dirac and electromagnetic
fields.

\subsection{Scalar field}

First we consider a scalar field $\varphi (x)$ with the mass $m$. Assuming a
general curvature coupling with the parameter $\xi $, the field equation
reads
\begin{equation}
\left( g^{\mu \rho }\nabla _{\mu }\nabla _{\rho }+m^{2}+\xi R\right) \varphi
(x)=0,  \label{fieldeq}
\end{equation}%
where $\nabla _{\mu }$ is the covariant derivative operator. The most
important special cases correspond to minimally ($\xi =0$) and conformally ($%
\xi =\xi _{D}=(D-1)/(4D)$) coupled fields. Let $\varphi _{\sigma }^{(\pm
)}(x)$ be a complete set of positive and negative energy mode functions
specified by the set of quantum numbers $\sigma $. In accordance with the
problem symmetry, they can be presented in the form%
\begin{equation}
\varphi _{\sigma }(x)=e^{i\mathbf{kx}\mp i\omega t}f(z),  \label{phisig}
\end{equation}%
where $\mathbf{x}=(x^{1},x^{2},\ldots ,x^{D-1})$, $\mathbf{k}%
=(k_{1},k_{2},\ldots ,k_{D-1})$, and $\mathbf{kx}=\sum_{l=1}^{D-1}k_{i}x^{i}$%
. Plugging into the field equation we get an ordinary differential equation
for the function $f(z)$:%
\begin{equation}
z^{D+1}\partial _{z}\left[ z^{1-D}\partial _{z}f(z)\right] +\left[ \lambda
^{2}z^{2}+\left( \xi D(D+1)-a^{2}m^{2}\right) \right] f(z)=0,  \label{Eqfsc}
\end{equation}%
where the energy $\omega $ is expressed in terms of $\lambda $ and $%
k^{2}=\sum_{i=1}^{D-1}k_{i}^{2}$ as $\omega =\sqrt{\lambda ^{2}+k^{2}}$. The
general solution of (\ref{Eqfsc}) is expressed in terms of the Bessel and
Neumann functions $J_{\nu }(\lambda z)$ and $Y_{\nu }(\lambda z)$:
\begin{equation}
f(z)=z^{D/2}\left[ c_{1}J_{\nu }(\lambda z)+c_{2}Y_{\nu }(\lambda z)\right] ,
\label{fz}
\end{equation}%
where $c_{1}$ and $c_{2}$ are constants and%
\begin{equation}
\nu =\sqrt{D^{2}/4-D(D+1)\xi +m^{2}a^{2}}.  \label{nu}
\end{equation}%
In order to have a stable vacuum we assume the values of the parameters in (%
\ref{nu}) for which $\nu \geqslant 0$ (see \cite{Brei82,Mezi85}). For a
conformally coupled massless field one has $\nu =1/2$ and the mode functions
(\ref{phisig}) with (\ref{fz}) are conformally related to the modes in the
Minkowski bulk. The scalar modes are normalized by the condition%
\begin{equation}
\int d^{D}x\,g^{00}\sqrt{|g|}\varphi _{\sigma }^{(s)}(x)\varphi _{\sigma
^{\prime }}^{(s^{\prime })\ast }(x)=\frac{\delta _{ss^{\prime }}}{2\omega }%
\delta _{\lambda \lambda ^{\prime }}\delta (\mathbf{k}-\mathbf{k}^{\prime }),
\label{Norm}
\end{equation}%
where $g$ is the determinant of the metric tensor. Here $\delta _{\lambda
\lambda ^{\prime }}$ is the Kronecker delta in the problems with discrete
eigenvalues of the quantum number $\lambda $ and $\delta _{\lambda \lambda
^{\prime }}=\delta (\lambda -\lambda ^{\prime })$ in the problems with
continuous spectrum for $\lambda $.

We are interested in the effects of a codimension one brane, localized at $%
z=z_{0}$, on the local properties of the scalar vacuum. The Robin boundary
condition will be imposed for the field operator on the brane:%
\begin{equation}
(\beta n^{\mu }\nabla _{\mu }+1)\varphi (x)=0,\;z=z_{0},  \label{Rob}
\end{equation}%
where $n^{\mu }$ is the inward pointing normal to the brane and $\beta $ is
a constant. The latter encodes the properties of the brane. In the special
cases $\beta =0$ and $\beta =\infty $ the condition (\ref{Rob}) is reduced
to the Dirichlet and Neumann boundary conditions, respectively. For the
normal in (\ref{Rob}) one has $n^{\mu }=\delta _{\mathrm{(J)}}\delta
_{D}^{\mu }z/a$, where J=L, $\delta _{\mathrm{(L)}}=-1$ in the L-region and
J=R, $\delta _{\mathrm{(R)}}=1$ in the R-region. In general, the values of
the constant $\beta $ could be different for those regions.

First let us consider the modes in the R-region. From the boundary condition
(\ref{Rob}) it follows that $c_{2}/c_{1}=-\bar{J}_{\nu }(\lambda z_{0})/\bar{%
Y}_{\nu }(\lambda z_{0})$ for the coefficients in (\ref{fz}). Here and
below, for a given function $F(x)$, we use the notation with the bar defined
in accordance with%
\begin{equation}
\bar{F}(x)=B_{0}xF^{\prime }(x)+A_{0}F(x),  \label{Fbar}
\end{equation}%
with the coefficients
\begin{equation}
A_{0}=1+\delta _{\mathrm{(J)}}\frac{D\beta }{2a},\quad B_{0}=\delta _{%
\mathrm{(J)}}\frac{\beta }{a}.  \label{A0}
\end{equation}%
The mode functions in the R-region obeying the boundary condition (\ref{Rob}%
) are presented as%
\begin{equation}
\varphi _{\mathrm{(R)}\sigma }^{(\pm )}(x)=C_{\mathrm{(R)}\sigma
}z^{D/2}g_{\nu }(\lambda z_{0},\lambda z)e^{i\mathbf{kx}\mp i\omega t},
\label{ModesR}
\end{equation}%
with the function
\begin{equation}
g_{\nu }(u,v)=J_{\nu }(v)\bar{Y}_{\nu }(u)-\bar{J}_{\nu }(u)Y_{\nu }(v).
\label{gnu}
\end{equation}%
The spectrum for $\lambda $ is continuous and from the normalization
condition (\ref{Norm}) with $\delta _{\lambda \lambda ^{\prime }}=\delta
(\lambda -\lambda ^{\prime })$ we get%
\begin{equation}
|C_{\mathrm{(R)}\sigma }|^{2}=\frac{\left( 2\pi a\right) ^{1-D}\lambda }{%
2\omega \left[ \bar{J}_{\nu }^{2}(\lambda z_{0})+\bar{Y}_{\nu }^{2}(\lambda
z_{0})\right] }.  \label{Csig}
\end{equation}%
The modes are specified by the set $\sigma =(\mathbf{k},\lambda )$ with $%
-\infty <k_{i}<+\infty $, $i=1,\ldots ,D-1$, and $0\leq \lambda <\infty $.
The analog of the mode functions (\ref{ModesR}) in the region between two
parallel branes on the AdS bulk has been considered in \cite{Saha05}.

Note that we could also have modes with purely imaginary $\lambda $, $%
\lambda =i|\lambda |$. For those modes $f(z)=c_{1}z^{D/2}K_{\nu }(|\lambda
|z)$, where $K_{\nu }(x)$ is the Macdonald function (the modes with the
modified Bessel function $I_{\nu }(|\lambda |z)$ are not normalizable). From
the boundary condition (\ref{Rob}) we get the equation for the allowed
values of $|\lambda |$: $\bar{K}_{\nu }(|\lambda |z_{0})=0$. The energy
corresponding to these modes is given by $\omega =\sqrt{k^{2}-|\lambda |^{2}}
$ and it becomes imaginary for $k<|\lambda |$. This leads to the instability
of the vacuum state. In order to exclude the unstable modes we restrict the
allowed values for the Robin coefficient in the region where the equation $%
\bar{K}_{\nu }(|\lambda |z_{0})=0$ has no roots. It can be seen that for
non-Dirichlet boundary conditions ($\beta \neq 0$) the corresponding
condition is expressed as $a/\beta <\nu -D/2$ (for more detailed discussion
in models with compact dimensions see \cite{Bell15}). For $a/\beta >\nu -D/2$
there is a single root. For the special value $\beta /a=1/(\nu -D/2)$ there
exists a mode with $\lambda =0$ and with the mode functions $\varphi _{(%
\mathrm{R})\sigma }^{(\pm )}(x)=C_{(\mathrm{R})\sigma }z^{D/2-\nu }e^{i%
\mathbf{kx}\mp ikt}$. For a minimally coupled massless scalar field one has $%
\nu =D/2$ and this special value corresponds to the Neumann boundary
condition. The corresponding mode functions do not depend on $z$.

In the L-region the integration over $z$ in (\ref{Norm}) goes over the
interval $z\in \lbrack 0,z_{0}]$. For the solutions (\ref{phisig}) with (\ref%
{fz}) and $c_{2}\neq 0$, the $z$-integral diverges at the lower limit $z=0$
in the range $\nu \geq 1$. Hence, in that range, from the normalizability
condition for the mode functions it follows that we should take $c_{2}=0$.
In the range $0\leq \nu <1$, the solution (\ref{fz}) with $c_{2}\neq 0$ is
normalizable and in order to uniquely define the mode functions an
additional boundary condition at the AdS boundary is required \cite%
{Brei82,Avis78}. The general class of allowed boundary conditions has been
discussed in \cite{Ishi04,Morl20}. In particular, they include the Dirichlet
and Neumann boundary conditions, the most frequently used in the literature.
Here, for the values of the parameters corresponding to the range $0\leq \nu
<1$ we will choose the Dirichlet condition which gives $c_{2}=0$. With this
choice, the mode functions in the L-region are specified as%
\begin{equation}
\varphi _{\mathrm{(L)}\sigma }^{(\pm )}(x)=C_{\mathrm{(L)}\sigma
}z^{D/2}J_{\nu }(\lambda z)e^{i\mathbf{kx}\mp i\omega t}.  \label{ModesL}
\end{equation}%
From the boundary condition (\ref{Rob}) on the brane we get the equation for
the eigenvalues of the quantum number $\lambda $:%
\begin{equation}
\bar{J}_{\nu }(\lambda z_{0})=0.  \label{Jzer}
\end{equation}%
If we denote by $x=\lambda _{\nu ,n}$ the positive zeros of the functions $%
\bar{J}_{\nu }(x)$, numerated by $n=1,2,\ldots $, then the eigenvalues are
given by $\lambda =\lambda _{\nu ,n}/z_{0}$. Note that the roots $\lambda
_{\nu ,n}$ do not depend on the location of the brane. From the
normalization condition (\ref{Norm}), with $\delta _{\lambda \lambda
^{\prime }}=\delta _{nn^{\prime }}$ and with the $z$-integration over $%
[0,z_{0}]$, one finds%
\begin{equation}
|C_{\sigma }|^{2}=\frac{\left( 2\pi a\right) ^{1-D}\lambda _{\nu ,n}T_{\nu
}(\lambda _{\nu ,n})}{z_{0}\sqrt{k^{2}z_{0}^{2}+\lambda _{\nu ,n}^{2}}},
\label{CR}
\end{equation}%
with the function $T_{\nu }(x)=x[x^{2}J_{\nu }^{\prime 2}(x)+(x^{2}-\nu
^{2})J_{\nu }^{2}(x)]^{-1}$.

Similar to the R-region, the stability condition for the vacuum state in the
L-region imposes restrictions to the allowed values of the Robin coefficient
$\beta $. That condition excludes the presence of purely imaginary roots $%
\lambda =i|\lambda |$ for the equation (\ref{Jzer}). It can be shown that
there are no such modes for $a/\beta <D/2+\nu $ and there is a single mode
for $a/\beta >D/2+\nu $. As seen, the stability condition in the L-region is
less restrictive than that for the R-region. In the special case $\beta
/a=1/\left( D/2+\nu \right) $ one has a mode with $\lambda =0$ with the mode
functions $\varphi _{(\mathrm{L})\sigma }^{(\pm )}(x)=C_{(\mathrm{L}%
)}z^{D/2+\nu }e^{i\mathbf{kx}\mp ikt}$.

\subsection{Dirac field}

Now we turn to a massive Dirac field $\psi (x)$. The dynamics of the field
is governed by the Dirac equation
\begin{equation}
\left[ i\gamma ^{\mu }\left( \partial _{\mu }+\Gamma _{\mu }\right) -m\right]
\psi (x)=0,  \label{FieldEq}
\end{equation}%
where $\Gamma _{\mu }$ is the spin connection. The curved spacetime Dirac
matrices $\gamma ^{\mu }$ are expressed in terms of the flat spacetime
matrices $\gamma ^{(b)}$ by the relation $\gamma ^{\mu }=e_{(b)}^{\mu
}\gamma ^{(b)}$, with $e_{(b)}^{\mu }$ being the tetrad fields. In the
coordinate system corresponding to (\ref{metric2}) the latter can be taken
as $e_{(b)}^{\mu }=(z/a)\delta _{b}^{\mu }$. For the components of the spin
connection this gives $\Gamma _{D}=0$ and $\Gamma _{i}=\eta _{il}\gamma
^{(D)}\gamma ^{(l)}/(2z)$ for $i=0,\ldots ,D-1$. In an irreducible
representation of the Clifford algebra the matrices $\gamma ^{(b)}$ are $%
N\times N$ matrices, where $N=2^{[(D+1)/2]}$ and the square brackets in the
exponent mean the integer part. Up to a similarity transformation, the
irreducible representation is unique in odd numbers of spatial dimension $D$%
. For even values of $D$ one has two inequivalent irreducible
representations. We use the flat spacetime gamma matrices in the
representation
\begin{equation}
\gamma ^{(0)}=\left(
\begin{array}{cc}
0 & \chi _{0} \\
\chi _{0}^{\dagger } & 0%
\end{array}%
\right) ,\;\gamma ^{(l)}=\left(
\begin{array}{cc}
0 & \chi _{l} \\
-\chi _{l}^{\dagger } & 0%
\end{array}%
\right) ,\;l=1,2,\ldots ,D-1,  \label{gam2}
\end{equation}%
and $\gamma ^{(D)}=si\,\mathrm{diag}(1,-1)$, where $s=\pm 1$. In odd spatial
dimensions the two values of the parameter $s$ correspond to two
inequivalent representations. The commutation relations for the $N/2\times
N/2$ matrices $\chi _{l}$ and for their hermitian conjugate matrices $\chi
_{l}^{\dagger }$ are obtained from those for the Dirac matrices $\gamma
^{(l)}$. They are reduced to the relations $\chi _{0}\chi _{l}^{\dagger
}=\chi _{l}\chi _{0}^{\dagger }$, $\chi _{0}^{\dagger }\chi _{l}=\chi
_{l}^{\dagger }\chi _{0}$, $\chi _{0}^{\dagger }\chi _{0}=1$ and $\chi
_{l}\chi _{i}^{\dagger }+\chi _{i}\chi _{l}^{\dagger }=2\delta _{li}$, $\chi
_{l}^{\dagger }\chi _{i}+\chi _{i}^{\dagger }\chi _{l}=2\delta _{li}$, for $%
l,i=1,2,\ldots ,D-1$. The representation (\ref{gam2}) for the construction
of the gamma matrices in AdS spacetime has been considered in \cite{Bell18}.
Another representation is taken in \cite{Eliz13}.

Assuming the dependence on the coordinates $(t,\mathbf{x})$ in the form $e^{i%
\mathbf{kx}\mp i\omega t}$ and decomposing the spinor $\psi (x)$ into the
upper and lower components, in the representation (\ref{gam2}) the
corresponding equations are separated. The dependence of those components on
the $z$-coordinate is expressed in terms of the function $c_{1}J_{ma\pm
s/2}(\lambda z)+c_{2}Y_{ma\pm s/2}(\lambda z)$, where the upper and lower
signs correspond to the upper and lower components. The coefficients are
determined by the normalization condition and by the boundary condition on
the brane at $z=z_{0}$. The positive and negative energy fermionic modes $%
\psi _{\sigma }^{(\pm )}(x)$, specified by the set of quantum numbers $%
\sigma $, are normalized by the condition
\begin{equation}
\int d^{D}x\,(a/z)^{D}\psi _{\sigma }^{(\pm )\dagger }\psi _{\sigma ^{\prime
}}^{(\pm )}=\delta _{\sigma \sigma ^{\prime }}.  \label{NormD}
\end{equation}%
As in the case of a scalar field, here $\delta _{\sigma \sigma ^{\prime }}$
is understood as the Dirac delta function for the continuous components of $%
\sigma $ and the Kronecker delta for discrete ones. On the brane at $z=z_{0}$
we impose the bag boundary condition
\begin{equation}
(1+i\gamma ^{\mu }n_{\mu })\psi (x)=0,\;z=z_{0},  \label{Bagbc}
\end{equation}%
where $n_{\mu }=-\delta _{\mathrm{(J)}}\delta _{\mu }^{D}a/z$, J=R,L, with $%
\delta _{\mathrm{(R)}}=-\delta _{\mathrm{(L)}}=1$ for the L- and R-regions.

In the R-region, $z_{0}\leq z<\infty $, from the boundary condition (\ref%
{Bagbc}), for the ratio of the coefficients in the linear combination of the
Bessel and Neumann functions one finds $c_{2}/c_{1}=-J_{ma+1/2}(\lambda
z_{0})/Y_{ma+1/2}(\lambda z_{0})$. The positive and negative energy mode
functions, obeying the boundary condition (\ref{Bagbc}), are expressed as%
\begin{eqnarray}
\psi _{\mathrm{(R)}\sigma }^{(+)}(x) &=&C_{\mathrm{(R)}\sigma }^{(+)}z^{%
\frac{D+1}{2}}e^{i\mathbf{kx}-i\omega t}\left(
\begin{array}{c}
\frac{\mathbf{k\chi }\chi _{0}^{\dagger }+i\lambda -\omega }{\omega }%
g_{ma+1/2,ma+s/2}(\lambda z_{0},\lambda z)w^{(\gamma )} \\
i\chi _{0}^{\dagger }\frac{\mathbf{k\chi }\chi _{0}^{\dagger }+i\lambda
+\omega }{\omega }g_{ma+1/2,ma-s/2}(\lambda z_{0},\lambda z)w^{(\gamma )}%
\end{array}%
\right) ,  \notag \\
\psi _{\mathrm{(R)}\sigma }^{(-)}(x) &=&C_{\mathrm{(R)}\sigma }^{(-)}z^{%
\frac{D+1}{2}}e^{i\mathbf{kx}+i\omega t}\left(
\begin{array}{c}
i\chi _{0}\frac{\mathbf{k\chi }^{\dagger }\chi _{0}-i\lambda +\omega }{%
\omega }g_{ma+1/2,ma+s/2}(\lambda z_{0},\lambda z)w^{(\gamma )} \\
\frac{\mathbf{k\chi }^{\dagger }\chi _{0}-i\lambda -\omega }{\omega }%
g_{ma+1/2,ma-s/2}(\lambda z_{0},\lambda z)w^{(\gamma )}%
\end{array}%
\right) ,  \label{ModesFR}
\end{eqnarray}%
where $\mathbf{k\chi }=\sum_{l=1}^{D-1}k_{l}\chi _{l}$ and%
\begin{equation}
g_{\mu ,\rho }(x,u)=J_{\mu }(x)Y_{\rho }(u)-J_{\rho }(u)Y_{\mu }(x).
\label{gemu}
\end{equation}%
In (\ref{ModesFR}), the one-column matrices $w^{(\gamma )}$, $\gamma =$ $%
1,\ldots ,N/2$, are introduced with the elements $w_{l}^{(\gamma )}=\delta
_{l\gamma }$ and having $N/2$ rows. The normalization constant is determined
from the condition (\ref{NormD}):%
\begin{equation}
\left\vert C_{\mathrm{(R)}\sigma }^{(\pm )}\right\vert ^{2}=\lambda \frac{%
\left[ J_{ma+1/2}^{2}(\lambda z_{0})+Y_{ma+1/2}^{2}(\lambda z_{0})\right]
^{-1}}{4\left( 2\pi \right) ^{D-1}a^{D}}.  \label{CRF}
\end{equation}%
The set of quantum numbers $\sigma $\ is specified as $\sigma =(\mathbf{k}%
,\lambda ,\gamma )$.

In the L-region and for $ma\geq 1/2$ from the normalizability condition we
get $c_{2}=0$. In the range of the mass corresponding to $ma<1/2$ the modes
with $c_{2}\neq 0$ are normalizable as well and an additional boundary
condition is required to uniquely define the mode functions. Here we
consider a special case that corresponds to the choice $c_{2}=0$ for all
values of the mass. The fermionic modes are given by the expressions
\begin{eqnarray}
\psi _{\mathrm{(L)}\sigma }^{(+)}(x) &=&C_{\mathrm{(L)}\sigma }^{(+)}z^{%
\frac{D+1}{2}}e^{i\mathbf{kx}-i\omega t}\left(
\begin{array}{c}
\frac{\mathbf{k\chi }\chi _{0}^{\dagger }+i\lambda -\omega }{\omega }%
J_{ma+s/2}(\lambda z)w^{(\gamma )} \\
i\chi _{0}^{\dagger }\frac{\mathbf{k\chi }\chi _{0}^{\dagger }+i\lambda
+\omega }{\omega }J_{ma-s/2}(\lambda z)w^{(\gamma )}%
\end{array}%
\right) ,  \notag \\
\psi _{\mathrm{(L)}\sigma }^{(-)}(x) &=&C_{\mathrm{(L)}\sigma }^{(-)}z^{%
\frac{D+1}{2}}e^{i\mathbf{kx}+i\omega t}\left(
\begin{array}{c}
i\chi _{0}\frac{\mathbf{k\chi }^{\dagger }\chi _{0}-i\lambda +\omega }{%
\omega }J_{ma+s/2}(\lambda z)w^{(\gamma )} \\
\frac{\mathbf{k\chi }^{\dagger }\chi _{0}-i\lambda -\omega }{\omega }%
J_{ma-s/2}(\lambda z)w^{(\gamma )}%
\end{array}%
\right) .  \label{ModesFL}
\end{eqnarray}%
The allowed values of $\lambda $ are determined from the boundary condition (%
\ref{Bagbc}) on the brane. They are roots of the equation
\begin{equation}
J_{ma-1/2}(\lambda z_{0})=0,  \label{FermLlamb}
\end{equation}%
and are expressed as $\lambda =\lambda _{ma-1/2,n}/z_{0}$. The normalization
coefficient is obtained from (\ref{NormD}) and is given by%
\begin{equation}
|C_{\mathrm{(L)}\sigma }^{(\pm )}|^{2}=\frac{J_{ma+1/2}^{-2}(\lambda
_{ma-1/2,n})}{2(2\pi )^{D-1}a^{D}z_{0}^{2}}.  \label{Cbpm}
\end{equation}%
Note that the eigenvalues for $\lambda $ are the same for both the
representations of the Clifford algebra.

\subsection{Electromagnetic field}

For the electromagnetic field with the vector potential $A_{\mu }(x)$, $\mu
=0,1,\ldots ,D$, the field equation reads%
\begin{equation}
\nabla _{\rho }F^{\mu \rho }=0,  \label{Meq}
\end{equation}%
where $F_{\mu \rho }=\partial _{\mu }A_{\rho }-\partial _{\rho }A_{\mu }$ is
the field strength tensor. In order to find a complete set of mode functions
$A_{\sigma \mu }(x)$ for the vector potential we impose the Lorenz condition
$\nabla _{\mu }A^{\mu }=0$ and an additional gauge condition $A^{D}=0$. For
the positive energy modes, presenting the dependence on the coordinates $%
(t,x^{1},\ldots ,x^{D-1})$ in the form $e^{i\mathbf{kx}-i\omega t}$, from
the field equation we can see that%
\begin{equation}
A_{\sigma \mu }(x)=\epsilon _{(\gamma )\mu }z^{D/2-1}\left[
c_{1}J_{D/2-1}(\lambda z)+c_{2}Y_{D/2-1}(\lambda z)\right] e^{i\mathbf{kx}%
-i\omega t}.  \label{ModesV}
\end{equation}%
Here, $\gamma =1,\ldots ,D-1$ correspond to different polarizations
specified by the polarization vector $\epsilon _{(\gamma )\mu }$. For the
latter one has the normalization condition $\eta ^{\mu \rho }\epsilon
_{(\gamma )\mu }\epsilon _{(\gamma ^{\prime })\rho }=-\delta _{\gamma \gamma
^{\prime }}$ and the constraints $\epsilon _{(\gamma )D}=0$ and $\eta ^{\mu
\rho }k_{\mu }\epsilon _{(\gamma )\rho }=0$. The latter two relations follow
from the gauge conditions. The modes (\ref{ModesV}) are normalized by the
condition
\begin{equation}
\int d^{D}x\sqrt{|g|}[A_{\sigma ^{\prime }\mu }^{\ast }\nabla ^{0}A_{\sigma
}^{\mu }-(\nabla ^{0}A_{\sigma ^{\prime }\mu }^{\ast })A_{\sigma }^{\mu
}]=4i\pi \delta _{\sigma \sigma ^{\prime }},  \label{NCV}
\end{equation}%
where the set of quantum numbers is given by $\sigma =(\mathbf{k},\lambda
,\gamma )$.

The coefficients $c_{1}$ and $c_{2}$ in (\ref{ModesV}) are determined from (%
\ref{NCV}) and from the boundary condition on the brane. We will consider
two types of gauge invariant constraints. The first condition is the analog
of the perfect conductor boundary condition in $D=3$ Maxwell electrodynamics
and is given by
\begin{equation}
n^{\mu _{1}}\,^{\ast }F_{\mu _{1}\cdots \mu _{D-1}}=0,\;z=z_{0},  \label{BC}
\end{equation}%
with $^{\ast }F_{\mu _{1}\cdots \mu _{D-1}}$ being the dual of the field
tensor and $n^{\mu }$ is the normal vector to the boundary. The second
boundary conditions is expressed as
\begin{equation}
n^{\mu }F_{\mu \rho }=0,\;z=z_{0}.  \label{BC2}
\end{equation}%
This type of condition has been used in quantum chromodynamics to confine
the gluons. In the gauge under consideration and for the mode functions (\ref%
{ModesV}) the boundary condition (\ref{BC}) yields $A_{\sigma
D}|_{z=z_{0}}=0 $, whereas the condition (\ref{BC2}) gives $\partial
_{D}A_{\sigma l}|_{z=z_{0}}=0$.

In the R-region, from the boundary condition on the brane we get $%
c_{2}/c_{1}=-J_{\nu }(\lambda z_{0})/Y_{\nu }(\lambda z_{0})$, where
\begin{equation}
\nu =\left\{
\begin{array}{cc}
D/2-1, & \text{for (\ref{BC}),} \\
D/2-2, & \text{for (\ref{BC2}).}%
\end{array}%
\right.  \label{nuV}
\end{equation}%
The corresponding mode functions for the vector potential are presented as%
\begin{equation}
A_{\sigma \mu }=C_{\mathrm{(R)}\sigma }\epsilon _{(\gamma )\mu
}z^{D/2-1}g_{\nu ,D/2-1}(\lambda z_{0},\lambda z)e^{i\mathbf{kx}-i\omega t},
\label{ModesVR}
\end{equation}%
where the function $g_{\mu ,\rho }(x,u)$ is given by (\ref{gemu}). The
normalization constant is found from (\ref{NCV}):%
\begin{equation}
C_{\mathrm{(R)}\sigma }^{2}\,=\lambda \frac{\left[ J_{\nu }^{2}(\lambda
z_{0})+Y_{\nu }^{2}(\lambda z_{0})\right] ^{-1}}{\left( 2\pi \right)
^{D-2}a^{D-3}\sqrt{k^{2}+\lambda ^{2}}},  \label{CRE}
\end{equation}%
with $0\leq \lambda <\infty $.

In the L-region from the normalizability of the mode functions it follows
that in (\ref{ModesV}) $c_{2}=0$ for $D\geq 4$. For $D=3$ an additional
condition on the AdS boundary is required in order to uniquely define the
modes. Here we consider a special case with $c_{2}=0$ for $D=3$ as well. For
the mode functions we obtain
\begin{equation}
A_{\sigma \mu }(x)=C_{\mathrm{(L)}\sigma }\epsilon _{(\gamma )\mu
}z^{D/2-1}J_{D/2-1}(\lambda z)e^{i\mathbf{kx}-i\omega t}.  \label{ModesVL}
\end{equation}%
The allowed values for $\lambda $ are determined by the boundary condition
on the brane and they are the roots of the equation%
\begin{equation}
J_{\nu }(\lambda z_{0})=0,  \label{JnuE}
\end{equation}%
with $\nu $ from (\ref{nuV}). Hence, we have $\lambda =\lambda _{\nu
,n}/z_{0}$. For the normalization coefficient one gets
\begin{equation}
C_{\mathrm{(L)}\sigma }^{2}=\frac{2\left( 2\pi \right) ^{2-D}a^{3-D}}{z_{0}%
\sqrt{k^{2}z_{0}^{2}+\lambda _{\nu ,n}^{2}}J_{\nu }^{\prime 2}(\lambda _{\nu
,n})}.  \label{CL}
\end{equation}%
The mode functions in the region between two branes have been recently
considered in \cite{Saha20}.

\subsection{Boundary conditions in $Z_{2}$-symmetric braneworlds}

An example of the $Z_{2}$-symmetric braneworld is provided by the
Randall-Sundrum model with a single brane (RSII model) formulated in
background of (4+1)-dimensional AdS spacetime (see \cite{Rand99a,Rand99b}
and the review \cite{Maar10} for the RSI and RSII models). For an arbitrary
number of spatial dimensions the line element is given by (\ref{metric})
with $e^{-2y/a}$ replaced by $e^{-2|y|/a}$. The regions $-\infty <y<0$ and $%
0<y<+\infty $ are identified by the $Z_{2}$-symmetry. The brane is located
at $y=0$. Hence, in the corresponding setup two copies of the R-region are
employed with $z_{0}=a$. The boundary conditions on the bulk fields at the
location of the brane are obtained by integrating the field equations about $%
y=0$ (see, e.g., the discussions in \cite%
{Flac01,Flac01c,Saha05,Bell18,Gher00,Chan05}).

For scalar fields even under the reflection with respect to the brane, the
Robin boundary condition is obtained with the coefficient $\beta
=-2/(c_{b}+4D\xi /a)$, where $c_{b}$ is the brane mass term. The latter
appears in the part of the action located on the brane, $S_{b}=-c_{b}\int
d^{D}xdy\sqrt{|g|}\delta (y)\varphi ^{2}/2$. For odd scalar fields the
Dirichlet boundary condition is obtained. For fermionic fields two types of
boundary conditions are obtained. The first one is reduced to the bag
boundary condition (\ref{Bagbc}) and the second one is obtained from (\ref%
{Bagbc}) by the change of the sign in the term containing the normal to the
brane. For vector fields even under the reflection with respect to the
brane, the boundary condition is reduced to (\ref{BC2}) and for odd fields
the condition (\ref{BC}) is obtained. For fermionic fields with the boundary
condition obtained from (\ref{Bagbc}) by the change of the sign in the
second term, the VEV of the energy-momentum tensor is evaluated in a way
similar to that we have demonstrated for the bag boundary condition. The
corresponding mode functions are obtained from (\ref{ModesFR}) by the
replacement $ma+1/2\rightarrow ma-1/2$ in the first index of the functions $%
g_{\mu ,\rho }(x,u)$, $g_{ma+1/2,ma+s/2}(\lambda z_{0},\lambda z)\rightarrow
g_{ma-1/2,ma+s/2}(\lambda z_{0},\lambda z)$, and in the expression (\ref{CRF}%
) for the normalization coefficient.

Hence, we conclude that the VEVs of the energy-momentum tensors for scalar,
Dirac and electromagnetic fields in $Z_{2}$-symmetric braneworlds with a
single brane are obtained from the results given below for the R-region by
an appropriate choice of the boundary conditions on the brane. The only
difference is that an additional factor $1/2$ should be added. The latter is
related to the presence of two copies of the R-region.

\section{Vacuum energy-momentum tensor}

\label{sec:EMT}

\subsection{General properties}

Having the mode functions for quantum fields we can investigate the VEVs of
the local characteristics of the vacuum state. Among the most important
characteristics is the VEV of the energy-momentum tensor. In particular, it
determines the distribution of the vacuum energy density and the forces
acting on the boundaries (the Casimir forces). For a free field $\Psi (x)$
(the only interaction is that with background gravitational field) the
operator of the energy-momentum tensor is a bilinear form in the field
operator:%
\begin{equation}
T_{\mu \rho }=T_{\mu \rho }\left\{ \Psi (x),\Psi (x)\right\} .  \label{Tmuop}
\end{equation}%
The corresponding expressions of the bilinear form in the right-hand side
for scalar, Dirac and vector fields can be found, for example, \cite%
{Grib94,Birr82}. Expanding the field operator in terms of a complete set $%
(\Psi _{\sigma }^{(+)}(x),\Psi _{\sigma }^{(-)}(x))$ of the positive and
negative energy mode functions, specified by quantum numbers $\sigma $,
using the commutation relations for the creation and annihilation operators
and the definition of the vacuum state $\left\vert 0\right\rangle $, for the
VEV of the energy-momentum tensor the following sum over the modes is
obtained:%
\begin{equation}
\left\langle 0\right\vert T_{\mu \rho }\left\vert 0\right\rangle \equiv
\left\langle T_{\mu \rho }\right\rangle =\frac{1}{2}\sum_{\sigma
}\sum_{s=\pm }T_{\mu \rho }\left\{ \Psi _{\sigma }^{(s)}(x),\Psi _{\sigma
}^{(s)\ast }(x)\right\} .  \label{Tmusum}
\end{equation}%
Here, $\sum_{\sigma }$ is understood as summation for discrete components of
$\sigma $ and as integration over the continuous components. The expression
in the right-hand side of (\ref{Tmusum}) is divergent and a regularization
procedure is required. The regularization can be made by point splitting, by
using the local zeta function technique or by introducing a cutoff function
\cite{Grib94}-\cite{Byts03}. In the presence of boundaries the VEV (\ref%
{Tmusum}) is decomposed into the boundary-free and boundary-induced
contributions. The structure of the divergences is uniquely determined by
the local geometric characteristics of the background spacetime. For points
away from boundaries the local geometry in the problems without and with
boundaries is the same and, hence, the divergences are the same as well.
From here it follows that at those points the boundary-induced contributions
in the VEVs of local observables are finite and the renormalization
procedure is the same as that in the boundary-free geometry. Consequently,
for points outside the boundaries the renormalization in (\ref{Tmusum}) is
reduced to that for the boundary-free part and the regularization
dependences may appear in that part only. For definiteness, we will assume
that a cutoff function is introduced in (\ref{Tmusum}) without writing it
explicitly. In \cite{Kota17}-\cite{Saha20},\cite{Saha06a,Saha06b,Saha05} the
point-splitting regularization technique has been used. The details of the
evaluation for the boundary-induced contributions at points outside the
boundaries do not depend on the specific regularization scheme.

In the geometry with a brane, we will decompose the vacuum energy-momentum
tensor into two contributions:%
\begin{equation}
\left\langle T_{\mu \rho }\right\rangle =\left\langle T_{\mu \rho
}\right\rangle _{0}+\left\langle T_{\mu \rho }\right\rangle _{\mathrm{b}},
\label{Tmudec}
\end{equation}%
where $\left\langle T_{\mu \rho }\right\rangle _{0}$ is the VEV in the
absence of the brane and the part $\left\langle T_{\mu \rho }\right\rangle _{%
\mathrm{b}}$ is induced by the brane. The VEV in the brane-free AdS geometry
has been widely discussed in the literature (for recent discussion see, for
example, \cite{Kent15,Ambr15}) and here we are mainly interested in the
brane-induced effects. From the maximal symmetry of the AdS spacetime it
follows that $\left\langle T_{\mu \rho }\right\rangle _{0}=\mathrm{const}%
\cdot g_{\mu \rho }$. On the basis of the symmetry of the problem with a
brane parallel to the AdS boundary, we expect that the VEV $\left\langle
T_{\mu \rho }\right\rangle _{\mathrm{b}}$ is diagonal. From the Lorentz
invariance in the subspace $(t,x^{1},\ldots ,x^{D-1})$ one concludes that
the stresses in the directions parallel to the brane are equal to the energy
density:
\begin{equation}
\left\langle T_{0}^{0}\right\rangle _{\mathrm{b}}=\left\langle
T_{1}^{1}\right\rangle _{\mathrm{b}}=\cdots =\left\langle
T_{D-1}^{D-1}\right\rangle _{\mathrm{b}}.  \label{Tmupar}
\end{equation}%
An additional relation between the components of the brane-induced VEV is
obtained from the covariant continuity equation $\nabla _{\mu }\left\langle
T_{\rho }^{\mu }\right\rangle _{\mathrm{b}}=0$. The latter is reduced to
\begin{equation}
z^{D+1}\partial _{z}\left( z^{-D}\left\langle T_{D}^{D}\right\rangle _{%
\mathrm{b}}\right) +D\left\langle T_{0}^{0}\right\rangle _{\mathrm{b}}=0.
\label{Conteq}
\end{equation}%
As it will be shown below, the brane-induced contribution in the VEVs depend
on the coordinate $z$ and on the location of the brane through the ratio $%
z/z_{0}$. This property is a consequence of the maximal symmetry of the AdS
spacetime and of the vacuum state we consider here. In addition, for points
outside the brane the following trace relations take place for scalar, Dirac
and electromagnetic fields:%
\begin{eqnarray}
\langle T_{\mu }^{\mu }\rangle _{\mathrm{b}}^{\mathrm{(s)}} &=&\left[ D(\xi
-\xi _{D})\nabla _{\mu }\nabla ^{\mu }+m^{2}\right] \langle \varphi
^{2}\rangle _{\mathrm{b}},  \notag \\
\langle T_{\mu }^{\mu }\rangle _{\mathrm{b}}^{\mathrm{(f)}} &=&m\left\langle
\bar{\psi }\psi \right\rangle _{\mathrm{b}},  \notag \\
\langle T_{\mu }^{\mu }\rangle _{\mathrm{b}}^{\mathrm{(v)}} &=&-\frac{D-3}{%
16\pi }\left\langle F_{\mu \rho }F^{\mu \rho }\right\rangle _{\mathrm{b}}.
\label{Trace}
\end{eqnarray}%
Here and below we use the superscripts (s),(f),(v) for the brane-induced
energy-momentum tensors in the cases of scalar, Dirac and electromagnetic
fields. On the right-hand sides of (\ref{Trace}) the subscript b stands for
the brane-induced contributions in the corresponding VEVs. For conformally
coupled fields (conformally coupled massless scalar field ($\xi =\xi _{D}$, $%
m=0$), massless Dirac field, the electromagnetic field in $D=3$ spatial
dimensions) the brane-induced energy-momentum tensor is traceless. For
points away from the brane the trace anomaly is contained in the part $%
\left\langle T_{\mu \rho }\right\rangle _{0}$ only (on trace anomalies for
different fields see, for example, \cite{Birr82}). The procedure for the
evaluation of the brane-induced contribution in the VEV\ of the
energy-momentum tensor follows similar steps for scalar, fermion and
electromagnetic fields and we illustrate it on the example of the fermionic
field.

The geometry given by (\ref{metric2}) is conformally flat and for
conformally coupled fields the problem under consideration is conformally
related to the corresponding problem in the Minkowski bulk. Denoting by $%
\left\langle T_{\mu }^{\rho }\right\rangle _{\mathrm{b}}^{\mathrm{(M)}}$ the
boundary-induced VEV in the Minkowskian problem, we have the following
relation%
\begin{equation}
\left\langle T_{\mu }^{\rho }\right\rangle _{\mathrm{b}}=(z/a)^{D+1}\left%
\langle T_{\mu }^{\rho }\right\rangle _{\mathrm{b}}^{\mathrm{(M)}}.
\label{ConfRel}
\end{equation}%
For the R-region, the problem with a brane in AdS bulk parallel to the AdS
boundary is conformally related to the problem in the Minkowski bulk with
the line element $ds_{\mathrm{M}}^{2}=\eta _{\mu \rho }dx^{\mu }dx^{\rho }$,
$x^{D}=z$, and with a single boundary at $z=z_{0}$. For the L-region the
Minkowskian counterpart contains two boundaries. The first one is located at
$z=z_{0}$ and is the conformal image of the brane and the second one is
located at $z=0$ and is the conformal image of the AdS boundary. The
boundary conditions on $z=0$ for fields in the Minkowskian problem are
related to the special boundary conditions we have imposed on the AdS
boundary.

\subsection{R-region}

We will describe the procedure for the evaluation of the brane-induced
energy-momentum tensor on the example of the Dirac field. The procedure for
scalar and electromagnetic fields follows similar steps.

\subsubsection{Dirac field}

By using the expression for the energy-momentum tensor of the Dirac field
and the corresponding mode functions from the previous section, the diagonal
components in the R-region are presented in the form%
\begin{eqnarray}
\langle T_{\mu }^{\rho }\rangle ^{\mathrm{(f)}} &=&\frac{\delta _{\mu
}^{\rho }Na^{-D-1}(z/z_{0})^{D+2}}{2(4\pi )^{(D-1)/2}\Gamma ((D-1)/2)}%
\int_{0}^{\infty }du\,u^{D-2}\mathbf{\,}  \notag \\
&&\times \int_{0}^{\infty }dx\frac{x}{\sqrt{x^{2}+u^{2}}}\frac{f_{\mathrm{(R)%
}}^{(\mu )}(x,xz/z_{0})}{J_{ma+1/2}^{2}(x)+Y_{ma+1/2}^{2}(x)},  \label{EMTs1}
\end{eqnarray}%
where
\begin{eqnarray}
f_{\mathrm{(R)}}^{(0)}(x,y) &=&-\left( x^{2}+u^{2}\right) \left[
g_{ma+1/2,ma+s/2}^{2}(x,y)+g_{ma+1/2,ma-s/2}^{2}(x,y)\right] ,  \notag \\
f_{\mathrm{(R)}}^{(D)}(x,y) &=&x^{2}\left[
g_{ma+1/2,ma+s/2}^{2}(x,y)+g_{ma+1/2,ma-s/2}^{2}(x,y)\right.  \notag \\
&&\left. -\frac{2ma}{y}g_{ma+1/2,ma+s/2}(x,y)g_{ma+1/2,ma-s/2}(x,y)\right] .
\label{fD}
\end{eqnarray}%
As seen from (\ref{fD}), the VEVs for $s=+1$ and $s=-1$ coincide. For the
separation of the brane-induced part we use the relation
\begin{equation}
\frac{g_{\mu ,\rho }(x,u)g_{\mu ,\rho ^{\prime }}(x,u)}{J_{\mu
}^{2}(x)+Y_{\mu }^{2}(x)}=J_{\rho }(u)J_{\rho ^{\prime }}(u)-\frac{1}{2}%
\sum_{s=1,2}\frac{J_{\mu }(x)}{H_{\mu }^{(s)}(x)}H_{\rho }^{(s)}(u)H_{\rho
^{\prime }}^{(s)}(u),  \label{Rel2}
\end{equation}%
where $H_{\rho }^{(s)}(u)$, $s=1,2$, are the Hankel functions. For the
separate terms in the integrand of (\ref{EMTs1}) one has $\mu =ma+1/2$, $%
\rho ,\rho ^{\prime }=ma\pm s/2$, and $u=xz/z_{0}$. The part in the VEV
coming from the first term in the right-hand side of (\ref{Rel2}) is the
vacuum energy-momentum tensor in the geometry where the brane is absent (the
part $\left\langle T_{\mu \rho }\right\rangle _{0}$ in (\ref{Tmudec})). In
the complex plane $x=re^{i\phi }$, with $r$ being the modulus of $x$, for $%
z>z_{0}$ and for large values of $r$ the $s=1$ term in (\ref{Rel2}) (with $%
u=xz/z_{0}$) is exponentially small in the quarter $0<\phi \leq \pi /2$ and
the $s=2$ term is exponentially small in the quarter $-\pi /2\leq \phi <0$.
On the basis of these properties, in the parts of the energy-momentum tensor
with the last term in (\ref{Rel2}) we rotate the integration contour over $x$
by the angle $\pi /2$ for $s=1$ terms and by the angle $-\pi /2$ for $s=-1$
terms. Introducing the modified Bessel functions $I_{\mu }(x)$ and $K_{\mu
}(x)$, we get
\begin{equation}
\langle T_{\mu }^{\rho }\rangle _{\mathrm{b}}^{\mathrm{(f)}}=-\frac{%
2^{-D}\delta _{\mu }^{\rho }N}{\pi ^{D/2}\Gamma (D/2)a^{D+1}}%
\int_{0}^{\infty }dx\,x^{D+1}\frac{I_{ma+1/2}(xz_{0}/z)}{K_{ma+1/2}(xz_{0}/z)%
}F_{\mathrm{(R)}}^{(\mu )}(x),  \label{EMTb1}
\end{equation}%
with the notations%
\begin{eqnarray}
F_{\mathrm{(R)}}^{(0)}(x) &=&\frac{1}{D}\left[
K_{ma+1/2}^{2}(x)-K_{ma-1/2}^{2}(x)\right] ,  \notag \\
F_{\mathrm{(R)}}^{(D)}(x) &=&K_{ma-1/2}^{2}(x)-K_{ma+1/2}^{2}(x)+\frac{2ma}{x%
}K_{ma+1/2}(x)K_{ma-1/2}(x).  \label{Snu}
\end{eqnarray}%
For massive fields both the energy density $\langle T_{0}^{0}\rangle _{%
\mathrm{b}}^{\mathrm{(f)}}$ and the normal stress $\langle T_{D}^{D}\rangle
_{\mathrm{b}}^{\mathrm{(f)}}$ are negative. As seen from (\ref{EMTb1}), the
brane-induced contribution depends on the coordinates $z$ and $z_{0}$
through the ratio $z/z_{0}=e^{(y-y_{0})/a}$, where $y_{0}=a\ln (z_{0}/a)$ is
the location of the brane in terms of the coordinate $y$. Note that $y-y_{0}$
is the proper distance from the brane. For a massless field the
brane-induced contribution vanishes, $\langle T_{\mu }^{\rho }\rangle _{%
\mathrm{b}}=0$. The massless fermionic field is conformally invariant and
this result could be directly obtained from the corresponding result for a
single boundary in the Minkowski bulk by using the relation (\ref{ConfRel}).

The Minkowskian result for a massive field is obtained from (\ref{EMTb1}) in
the limit $a\rightarrow \infty $ for fixed $y$ (see (\ref{metric})). From
the relation $z=ae^{y/a}$ it follows that $z\approx a+y$ and the values for
the coordinate $z$ are large. By using the uniform asymptotic expansions for
the modified Bessel functions for large values of the order, the leading
term in the expansion of the components with $\mu =0,1,\ldots ,D-1$
coincides with the VEV for a planar boundary in the Minkowski bulk:%
\begin{equation}
\langle T_{\mu }^{\rho }\rangle _{\mathrm{b}}^{\mathrm{(Mf)}}=-\frac{%
2^{-D}\delta _{\mu }^{\rho }Nm}{\pi ^{D/2}D\Gamma (D/2)}\int_{m}^{\infty
}dx\,(x^{2}-m^{2})^{D/2}\frac{e^{-2x(y-y_{0})}}{x+m},  \label{EMTRD}
\end{equation}%
and the normal stress vanishes, $\langle T_{D}^{D}\rangle _{\mathrm{b}}^{%
\mathrm{(Mf)}}=0$.

The general formula (\ref{EMTb1}) for the brane-induced contribution in the
energy-momentum tensor is simplified in the region near the brane and at
large distances from it (near-horizon limit). For large values $x$ the
integrand in (\ref{EMTb1}) behaves as $x^{D-1}e^{-2x(1-z_{0}/z)}$ and the
integral is divergent for points on the brane. These surface divergences are
well-known in quantum field theory with boundaries and have been widely
discussed in the literature for different geometries of boundaries (see, for
instance, \cite{Most97}-\cite{Casi11}). Near the brane, assuming that $%
z/z_{0}-1\ll 1$, the contribution of large values of $x$ dominate in the
integral and by making use of the corresponding asymptotic formulas for the
modified Bessel functions, in the leading order we obtain%
\begin{equation}
\langle T_{0}^{0}\rangle _{\mathrm{b}}^{\mathrm{(f)}}\approx -\frac{Nm\Gamma
((D+1)/2)}{(4\pi )^{(D+1)/2}D\left( y-y_{0}\right) ^{D}},\;\langle
T_{D}^{D}\rangle _{\mathrm{b}}^{\mathrm{(f)}}\approx -\frac{DNm\Gamma
((D-1)/2)}{2(4\pi )^{(D+1)/2}Da(y-y_{0})^{D-1}}.  \label{EMTnear}
\end{equation}%
In terms of the coordinate $y$, these asymptotics are valid under the
conditions $y-y_{0}\ll a,m^{-1}$. The leading terms for the energy density
and for the stresses parallel to the brane coincide with those for a
boundary in the Minkowski bulk. This is related to the fact that near the
brane the dominant contribution to the VEV comes from the vacuum
fluctuations with wavelengths smaller than the curvatures radius and the
influence of the gravitational field on those fluctuations is weak. The
effects of gravity are essential at distances from the brane larger than the
curvature radius, $z/z_{0}\gg 1$ or $y-y_{0}\gg a$. In that region the
leading term in the asymptotic expansion for the energy density is expressed
as
\begin{eqnarray}
\langle T_{0}^{0}\rangle _{\mathrm{b}}^{\mathrm{(f)}} &\approx &-Nm\frac{%
\exp [-\left( 2m+1/a\right) (y-y_{0})]}{2^{D+2ma+1}\pi ^{(D-1)/2}a^{D}}
\notag \\
&&\times \frac{\Gamma (ma+(D+3)/2)\Gamma (2ma+D/2+1)}{(2ma+1)\Gamma
^{2}(ma+1/2)\Gamma (ma+D/2+2)},  \label{EMTLarge}
\end{eqnarray}%
and for the normal stress we get $\langle T_{D}^{D}\rangle _{\mathrm{b}}^{%
\mathrm{(f)}}\approx D\langle T_{0}^{0}\rangle _{\mathrm{b}}^{\mathrm{(f)}%
}/\left( D+2ma+1\right) $. Note that for a boundary in the Minkowski bulk
and for $y-y_{0}\gg 1/m$ the VEV decays like $e^{-2m(y-y_{0})}$. In the AdS
spacetime, the decay of the boundary-induced contribution, as a function of
the proper distance from the boundary, is stronger.

\begin{figure}[tbph]
\begin{center}
\begin{tabular}{cc}
\epsfig{figure=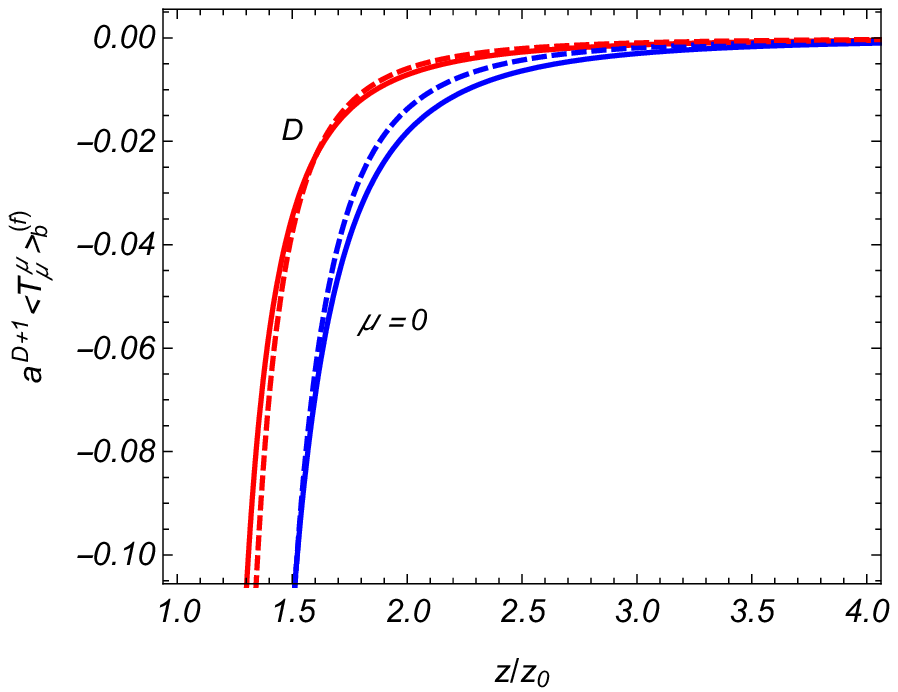,width=7.cm,height=5.5cm} & \quad %
\epsfig{figure=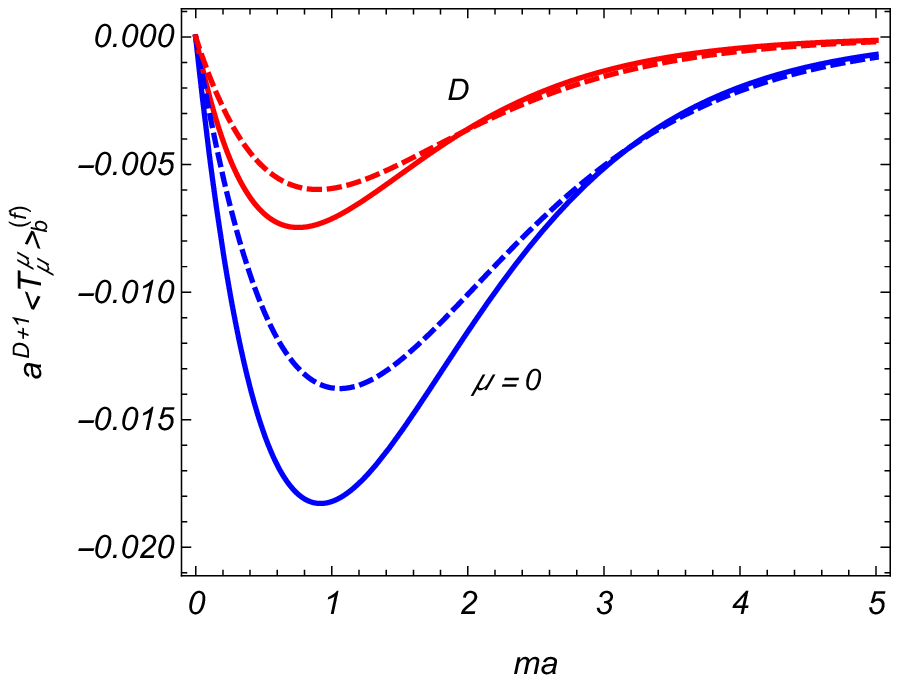,width=7.cm,height=5.5cm}%
\end{tabular}%
\end{center}
\caption{The brane-induced contributions in the VEVs of the energy density
and the normal stress as functions of $z/z_{0}$ (left panel) and of $ma$
(right panel). For the left panel we have taken $ma=1$ and for the right
panel $z/z_{0}=2$. The full and dashed curves correspond to $D=3$ and $D=4$.}
\label{fig1}
\end{figure}

The left panel of figure \ref{fig1} presents the dependence of the
brane-induced contributions to the energy density and the normal stress (in
units of $1/a^{D+1}$), $\langle T_{\mu }^{\mu }\rangle _{\mathrm{b}}^{%
\mathrm{(f)}}$ (no summation over $\mu $), $\mu =0,D$, on the ratio $z/z_{0}$
for fixed value $ma=1$. On the right panel we have displayed the
brane-induced parts as functions of the field mass for fixed $z/z_{0}=2$.
For both panels the full and dashed curves correspond to $D=3$ and $D=4$,
respectively.

\subsubsection{Scalar field}

In a similar way, for the brane-induced contribution in the case of a scalar
field with the boundary condition (\ref{Rob}) one obtains%
\begin{equation}
\langle T_{\mu }^{\rho }\rangle _{\mathrm{b}}^{\mathrm{(s)}}=-\frac{%
2^{-D}\delta _{\mu }^{\rho }}{\pi ^{D/2}\Gamma (D/2)a^{D+1}}\int_{0}^{\infty
}dx\,x^{D+1}\frac{\bar{I}_{\nu }(xz_{0}/z)}{\bar{K}_{\nu }(xz_{0}/z)}S^{(\mu
)}\left[ K_{\nu }(x)\right] ,  \label{Tik1plnew}
\end{equation}%
where the notation with the bar is defined as (\ref{Fbar}) and
\begin{eqnarray}
S^{(0)}[g(x)] &=&-\xi _{1}\left[ g^{\prime 2}(x)+\frac{D+4\xi /\xi _{1}}{x}%
g(x)g^{\prime }(x)+\left( 1+\frac{2}{D\xi _{1}}+\frac{\nu ^{2}}{x^{2}}%
\right) g^{2}(x)\right] ,  \notag \\
S^{(D)}[g(x)] &=&-g^{\prime 2}(x)+\frac{D\xi _{1}}{x}g(x)g^{\prime
}(x)+\left( 1+\frac{2m^{2}a^{2}-\nu ^{2}}{x^{2}}\right) g^{2}(x),
\label{SDnew}
\end{eqnarray}%
with%
\begin{equation}
\xi _{1}=4\xi -1.  \label{ksi1}
\end{equation}%
For a conformally coupled massless scalar field the brane-induced
contribution (\ref{Tik1plnew}) vanishes. In the limit $a\rightarrow \infty $%
, with fixed $y$ and $y_{0}$, from (\ref{Tik1plnew}) we get the vacuum
energy-momentum tensor for a Robin boundary in the Minkowski bulk:
\begin{equation}
\langle T_{\mu }^{\rho }\rangle _{\mathrm{b}}^{\mathrm{(Ms)}}=\frac{%
2^{-D}\delta _{\mu }^{\rho }}{\pi ^{D/2}\Gamma (D/2)}\int_{m}^{\infty
}dx\,\left( x^{2}-m^{2}\right) ^{D/2-1}\left[ \frac{m^{2}}{D}-4\left( \xi
-\xi _{D}\right) x^{2}\right] \frac{\beta x+1}{\beta x-1}e^{-2x(y-y_{0})},
\label{TmuM}
\end{equation}%
for $\mu =0,1,2,\ldots ,D-1$ and $\langle T_{D}^{D}\rangle _{\mathrm{b}}^{%
\mathrm{(Ms)}}=0$.

For points near the brane, $y-y_{0}\ll a,m^{-1},|\beta |$, the leading term
in the asymptotic expansion of the energy density is given by the expression%
\begin{equation}
\langle T_{0}^{0}\rangle _{\mathrm{b}}^{\mathrm{(s)}}\approx \pm \frac{%
D\Gamma \left( (D+1)/2\right) (\xi -\xi _{D})}{2^{D}\pi
^{(D+1)/2}(y-y_{0})^{D+1}},  \label{T00nearS}
\end{equation}%
where the upper and lower signs correspond to $\beta =0$ (Dirichlet boundary
condition) and $\beta \neq 0$ (non-Dirichlet boundary conditions),
respectively. The leading term in (\ref{T00nearS}) coincides with that for
the Minkwoski bulk. For the normal stress we find $\langle T_{D}^{D}\rangle
_{\mathrm{b}}^{\mathrm{(s)}}\approx \langle T_{0}^{0}\rangle _{\mathrm{b}}^{%
\mathrm{(s)}}(y-y_{0})/a$. In the case of a conformally coupled field, the
leading terms vanish and the divergences on the brane are weaker. At large
distances from the brane and for $\nu >0$, the energy density is
approximated as%
\begin{eqnarray}
\langle T_{0}^{0}\rangle _{\mathrm{b}}^{\mathrm{(s)}} &\approx &\frac{(2\nu
-1)e^{-2\nu (y-y_{0})/a}}{2^{2\nu +D-1}\pi ^{(D-1)/2}a^{D+1}}\frac{%
A_{0}+B_{0}\nu }{A_{0}-B_{0}\nu }\left( 4\xi -\frac{D+2\nu }{D+2\nu +1}%
\right)  \notag \\
&&\times \,\frac{\Gamma (D/2+\nu +1)\Gamma (D/2+2\nu )}{\Gamma (\nu
+1)\Gamma (D/2+1/2+\nu )}.  \label{T00farS}
\end{eqnarray}%
For the normal stress we get $\langle T_{D}^{D}\rangle _{\mathrm{b}}^{%
\mathrm{(s)}}\approx D\langle T_{0}^{0}\rangle _{\mathrm{b}}^{\mathrm{(s)}%
}/(D+2\nu )$. Note that for $\nu >0$ the decay of the brane-induced VEV\ is
exponential for both massless and massive fields. In the Minkowski bulk, the
boundary-induced contribution decays as $e^{-2m(y-y_{0})}$ for massive
fields and as power-law $(y-y_{0})^{-D-1}$ for non-conformally coupled
massless fields.

\begin{figure}[tbph]
\begin{center}
\begin{tabular}{cc}
\epsfig{figure=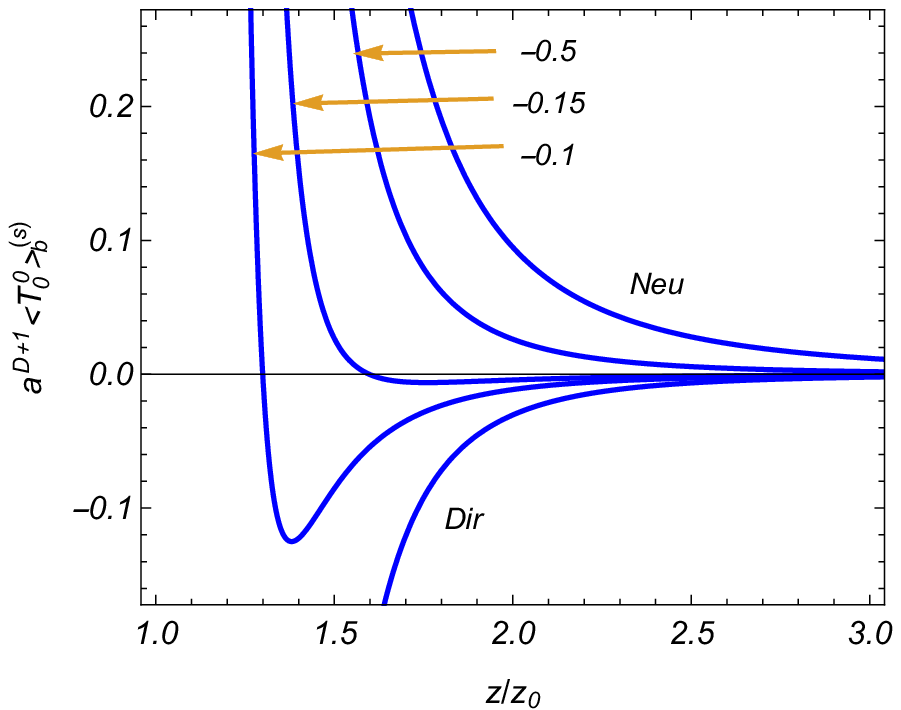,width=7.cm,height=5.5cm} & \quad %
\epsfig{figure=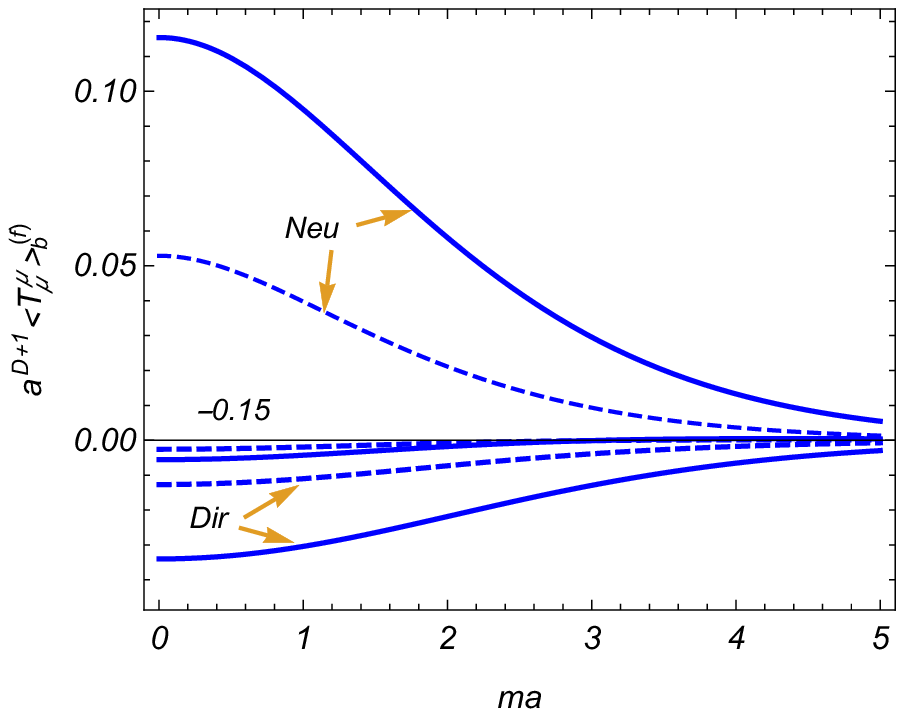,width=7.cm,height=5.5cm}%
\end{tabular}%
\end{center}
\caption{The brane-induced contributions in the VEVs of the energy density
and the normal stress for $D=4$ minimally coupled scalar field. For the left
panel $ma=1$ and for the right panel $z/z_{0}=2$. The full and dashed curves
on the right panel correspond to the energy density and normal stress,
respectively (for the boundary conditions chosen see the text). }
\label{fig2}
\end{figure}

On the left panel of figure \ref{fig2} we display the brane-induced energy
density in the R-region as a function of $z/z_{0}$ for $D=4$ minimally
coupled scalar field with $ma=1$. The graphs are plotted for the Dirichlet
and Neumann boundary conditions and for the Robin boundary conditions with $%
\beta /a=-0.1,-0.15,-0.5$. The graphs for the Robin boundary conditions are
located between the curves corresponding to the Dirichlet and Neumann
conditions and for them $\langle T_{0}^{0}\rangle _{\mathrm{b}}^{\mathrm{(s)}%
}$ increases with increasing $|\beta |$. The normal stress displays a
similar behavior as a function of $z/z_{0}$. The right panel in figure \ref%
{fig2} presents the brane-induced contributions to the energy density ($\mu
=0$, full curves) and normal stress ($\mu =D$, dashed curves) for fixed $%
z/z_{0}=2$ as functions of $ma$ for $D=4$ minimally coupled field. The
graphs are plotted for the Dirichlet and Neumann boundary conditions and for
the Robin boundary condition with $\beta /a=-0.15$. In the latter case the
energy density changes the sign at $ma\approx 3.13$ and the normal stress
changes the sign at $ma\approx 3.34$.

\begin{figure}[tbph]
\begin{center}
\begin{tabular}{cc}
\epsfig{figure=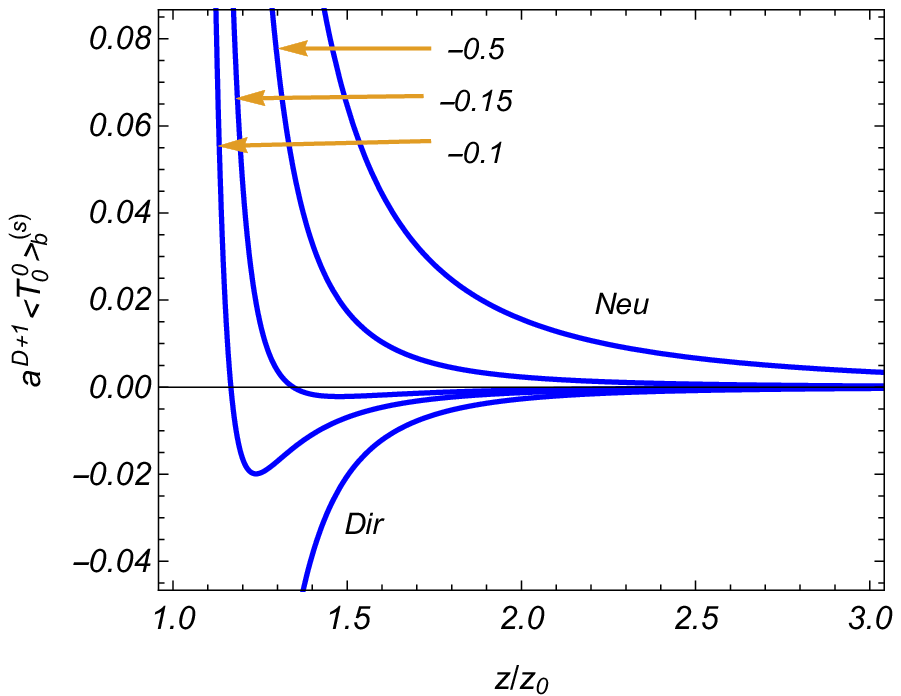,width=7.cm,height=5.5cm} & \quad %
\epsfig{figure=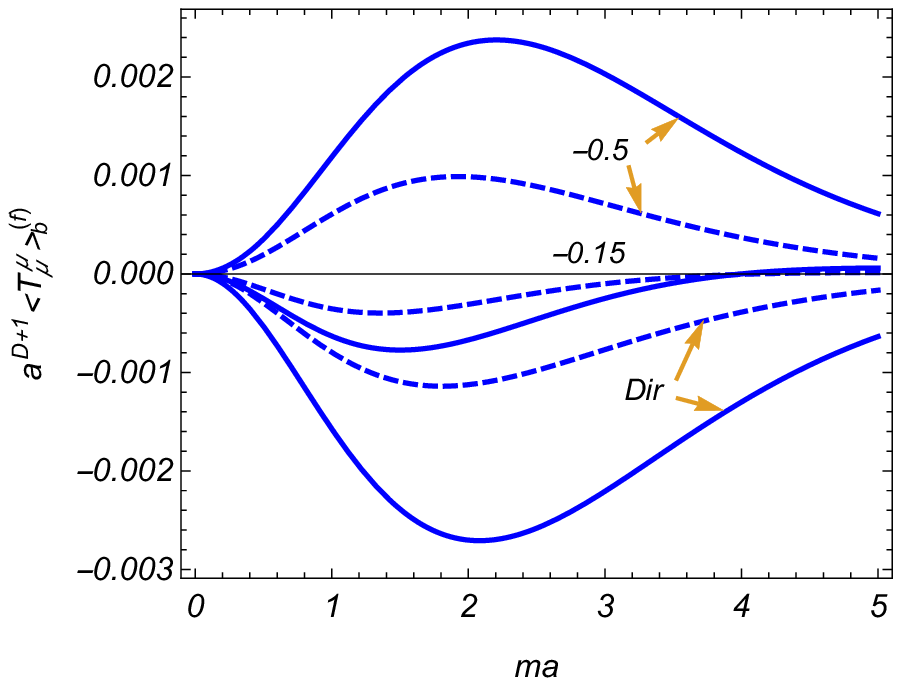,width=7.cm,height=5.5cm}%
\end{tabular}%
\end{center}
\caption{The same as in figure \protect\ref{fig2} for $D=4$ conformally
coupled scalar field. The left and right panels are plotted for $ma=2$ and $%
z/z_{0}=2$, respectively.}
\label{fig3}
\end{figure}

Figure \ref{fig3} presents similar results for $D=4$ conformally coupled
field ($\xi =3/16$). Note that in this case the stability condition $a/\beta
<\nu -D/2$ on the Robin coefficient is stronger. For the left panel we have
taken $ma=2$. With this value, for all the boundary conditions with $\beta
\leq 0$ the vacuum is stable. The right panel is plotted for $z/z_{0}=2$.
There is a range for the values of $ma$ in which the Neumann boundary
condition for a conformally coupled field leads to the vacuum instability.

An interesting feature that is seen from the graphs is the sign change and
nonmonotonicity of the brane-induced energy density (as a function of the
distance from the brane) for the Robin boundary condition. For a boundary in
the Minkowski bulk such behavior is easily obtained from the integral
representation (\ref{TmuM}). Indeed, for points near the boundary the
dominant contribution to the integral in (\ref{TmuM}) comes from large
values of $x$ and for $\beta \neq 0$ the leading term coincides with that
for the Neumann boundary condition. It is given by (\ref{T00nearS}) with the
lower sign. At large distances from the boundary the contribution of the
integration range near the lower limit dominates and to the leading order
one gets%
\begin{equation}
\langle T_{0}^{0}\rangle _{\mathrm{b}}^{\mathrm{(Ms)}}\approx \frac{\left(
1-4\xi \right) m^{D/2+1}e^{-2m(y-y_{0})}}{2^{D+1}\pi ^{D/2}(y-y_{0})^{D/2}}%
\frac{\beta m+1}{\beta m-1}.  \label{T00Mfar}
\end{equation}%
For a given curvature coupling parameter, depending on the Robin
coefficient, the energy density corresponding to (\ref{T00Mfar}) can be
either negative or positive. In particular, for a minimally coupled field we
see that the energy density is positive near the boundary for $\beta \neq 0$
and negative at large distances in the range $|\beta |<1/m$. This means that
the energy density changes the sign at some intermediate value of the
distance from the boundary. Similar analysis can be provided for a brane in
AdS bulk. Near the brane the leading term in the asymptotic expansion is the
same and the energy density is positive for non-Dirichlet boundary
conditions. At large distances, the asymptotic of the brane-induced VEV is
given by (\ref{T00farS}). For massive minimally and conformally coupled
fields one has $\nu >1/2$ and the sign of the energy density in (\ref%
{T00farS}) is determined by the sign of the fraction containing the Robin
coefficient. In particular, at large distances the energy density is
negative for $\left\vert a/\beta +D/2\right\vert >\nu $. Under this
constraint and for non-Dirichlet boundary conditions the brane-induced
energy density is positive near the brane and negative at large distances.
Examples of this kind of behavior are given in figures \ref{fig2} and \ref%
{fig3}.

\subsubsection{Electromagnetic field}

For the electromagnetic field with the boundary conditions (\ref{BC}) and (%
\ref{BC2}) on the brane, the brane-induced VEV has the form%
\begin{equation}
\langle T_{\mu }^{\rho }\rangle _{\mathrm{b}}^{\mathrm{(v)}}=-\frac{\delta
_{\nu }\delta _{\mu }^{\rho }\left( D-1\right) }{2^{D}\pi ^{D/2}\Gamma
(D/2)a^{D+1}}\int_{0}^{\infty }dx\,x^{D+1}\frac{I_{\nu }(xz_{0}/z)}{K_{\nu
}(xz_{0}/z)}V_{\mathrm{(R)}}^{(\mu )}(x),  \label{TmuVR}
\end{equation}%
where $\delta _{\nu }=1$ for $\nu =D/2-1$ (boundary condition (\ref{BC})), $%
\delta _{\nu }=-1$ for $\nu =D/2-2$ (boundary condition (\ref{BC2})), and
\begin{eqnarray}
V_{\mathrm{(R)}}^{(0)}(x) &=&\left( 1-\frac{4}{D}\right)
K_{D/2-1}^{2}(x)+\left( 1-\frac{2}{D}\right) K_{D/2-2}^{2}(x),  \notag \\
V_{\mathrm{(R)}}^{(D)}(x) &=&K_{D/2-1}^{2}(x)-K_{D/2-2}^{2}(x).  \label{VD}
\end{eqnarray}%
Note that the VEV has different signs for the boundary conditions (\ref{BC})
and (\ref{BC2}). For $D=3$ the electromagnetic field is conformally
invariant and the VEV (\ref{TmuVR}) is zero. One has (no summation over $\mu
$) $\langle T_{\mu }^{\mu }\rangle _{\mathrm{b}}^{\mathrm{(v)}}<0$ for the
boundary condition (\ref{BC}) and $\langle T_{\mu }^{\mu }\rangle _{\mathrm{b%
}}^{\mathrm{(v)}}>0$ for the condition (\ref{BC}). The VEVs for a plate in
the Minkowski bulk are obtained from (\ref{TmuVR}) in the limit $%
a\rightarrow \infty $:
\begin{equation}
\langle T_{\mu }^{\rho }\rangle _{\mathrm{b}}^{\mathrm{(Mv)}}=\mp \delta
_{\mu }^{\rho }\frac{(D-1)(D-3)\Gamma ((D+1)/2)}{2\left( 4\pi \right)
^{(D+1)/2}\left( y-y_{0}\right) ^{D+1}},  \label{TmuVM}
\end{equation}%
for $\mu =0,1,2,\ldots ,D-1$ and $\langle T_{D}^{D}\rangle _{\mathrm{b}}^{%
\mathrm{(Mv)}}=0$. The upper and lower signs in (\ref{TmuVM}) correspond to
the conditions (\ref{BC}) and (\ref{BC2}), respectively.

For $D>3$ the VEV (\ref{TmuVR}) diverges on the brane. The leading terms in
the asymptotic expansions for the energy density and for the normal stress
over the distance from the brane are expressed as
\begin{equation}
\langle T_{0}^{0}\rangle _{\mathrm{b}}^{\mathrm{(v)}}\approx -\delta _{\nu }%
\frac{\left( D-1\right) \left( D-3\right) \Gamma ((D+1)/2)}{2\left( 4\pi
\right) ^{(D+1)/2}\left( y-y_{0}\right) ^{D+1}},\;\langle T_{D}^{D}\rangle _{%
\mathrm{b}}^{\mathrm{(v)}}\approx \frac{y-y_{j}}{a}\langle T_{0}^{0}\rangle .
\label{T00near}
\end{equation}%
The leading term for the energy density coincides with the exact result for
a boundary in the Minkowski bulk, given by (\ref{TmuVM}). At large distances
from the brane, assuming that $z_{0}/z\ll 1$, for $\nu >0$ to the leading
order one gets%
\begin{equation}
\langle T_{\mu }^{\rho }\rangle _{\mathrm{b}}^{\mathrm{(v)}}\approx -\delta
_{\mu }^{\rho }\frac{2^{1-D-2\nu }\delta _{\nu }\left( D-1\right) e^{-2\nu
(y-y_{0})}}{\pi ^{D/2}\Gamma (D/2)\nu \Gamma ^{2}(\nu )a^{D+1}}%
\int_{0}^{\infty }dx\,x^{D+2\nu +1}V_{\mathrm{(R)}}^{(\mu )}(x),
\label{TmuVlarge}
\end{equation}%
where the integral is expressed in terms of the product of the gamma
functions (see \cite{Prud86}). For $D=4$ and for the boundary condition (\ref%
{BC2}) one has $\nu >0$ and at large distances the brane-induced part in the
energy-momentum tensor falls as $1/(y-y_{0})$.

\begin{figure}[tbph]
\begin{center}
\begin{tabular}{cc}
\epsfig{figure=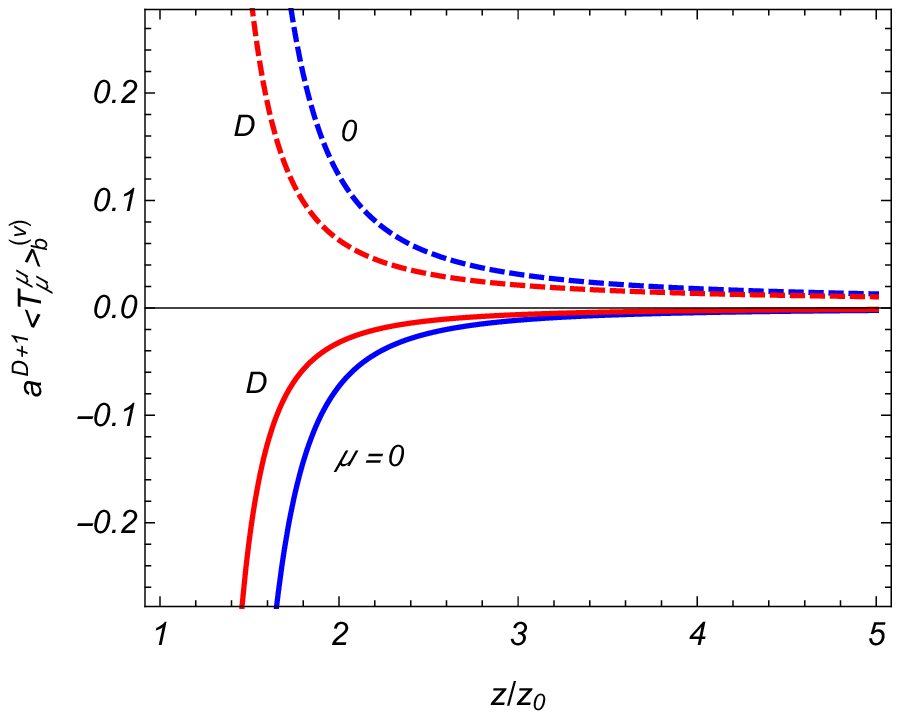,width=7.cm,height=5.5cm} & \quad %
\epsfig{figure=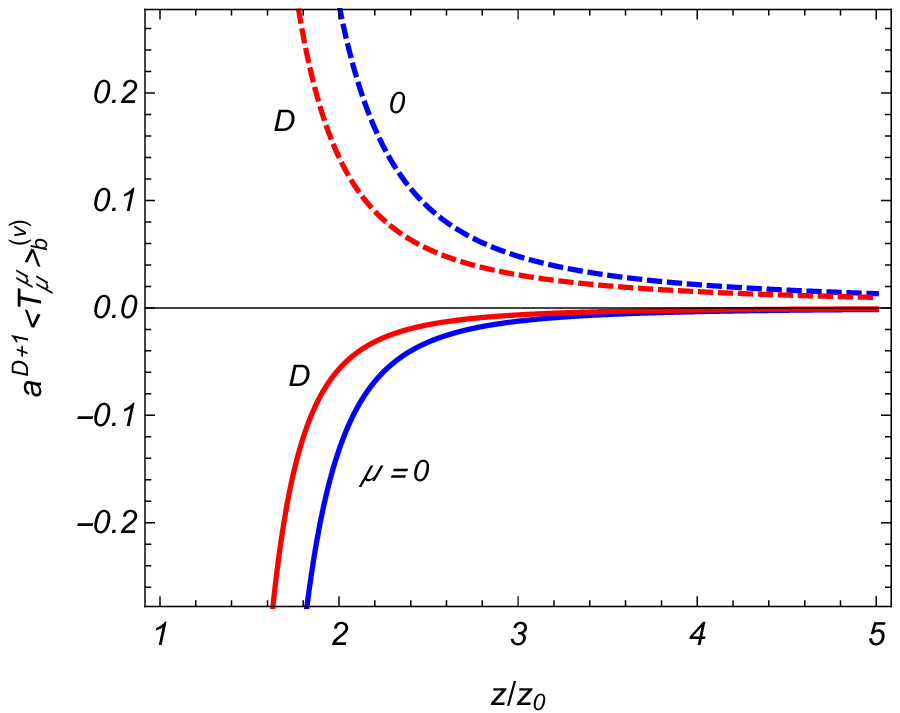,width=7.cm,height=5.5cm}%
\end{tabular}%
\end{center}
\caption{The brane-induced VEVs of the energy density and the normal stress
as functions of $z/z_{0}$ for $D=4$ (left panel) and $D=5$ (right panel)
electromagnetic fields. The full and dashed curves correspond to the
conditions (\protect\ref{BC}) and (\protect\ref{BC2}).}
\label{fig4}
\end{figure}

In figure \ref{fig4} we present the brane-induced energy density and the
normal stress versus $z/z_{0}$ for $D=4$ (left panel) and $D=5$ (right
panel) electromagnetic fields. The full and dashed curves correspond to the
boundary conditions (\ref{BC}) and (\ref{BC2}), respectively, and near the
curves the value of the index $\mu $ is given for $\langle T_{\mu }^{\mu
}\rangle _{\mathrm{b}}^{\mathrm{(v)}}$ (no summation over $\mu $) from (\ref%
{TmuVR}).

\subsection{L-region}

In the L-region the eigenvalues for the quantum number $\lambda $ are
discrete. Again, we illustrate the evaluation procedure for the example of
the Dirac field. The VEVs for scalar and electromagnetic fields are
evaluated in a similar way.

\subsubsection{Dirac field}

By using the mode functions (\ref{ModesFL}), the mode sum for the
energy-momentum tensor is presented in the form%
\begin{equation}
\langle T_{\mu }^{\rho }\rangle ^{\mathrm{(f)}}=-\frac{\delta _{\mu }^{\rho
}(4\pi )^{(1-D)/2}Nz^{D}}{\Gamma \left( (D-1)/2\right) a^{D+1}z_{0}}%
\int_{0}^{\infty }dk\,k^{D-2}\sum_{n=1}^{\infty }\frac{f_{\mathrm{(L)}%
}^{(\mu )}(\lambda _{ma-1/2,n}z/z_{0})}{\sqrt{k^{2}z_{0}^{2}+\lambda
_{ma-1/2,n}^{2}}J_{ma+1/2}^{2}(\lambda _{ma-1/2,n})},  \label{TmufL}
\end{equation}%
with the functions%
\begin{eqnarray}
f_{\mathrm{(L)}}^{(0)}(x) &=&-(k^{2}z^{2}+x^{2})\left[
J_{ma+s/2}^{2}(x)+J_{ma-s/2}^{2}(x)\right] ,  \notag \\
f_{\mathrm{(L)}}^{(D)}(x) &=&x^{2}\left[ J_{ma+s/2}^{2}(x)+J_{ma-s/2}^{2}(x)-%
\frac{2ma}{x}J_{ma+s/2}(x)J_{ma-s/2}(x)\right] .  \label{fDL}
\end{eqnarray}%
By taking into account that the eigenvalues $\lambda _{ma-1/2,n}$ do not
depend on $s$, we conclude that, similar to the R-region, the VEV of the
energy-momentum tensor is the same for both inequivalent representations of
the Clifford algebra.

For the summation of the series over the roots $\lambda _{ma-1/2,n}$ we use
the Abel-Plana type formula \cite{Saha87,Saha08}
\begin{eqnarray}
\sum_{n=1}^{\infty }\frac{f(\lambda _{ma-1/2,n})/\lambda _{ma-1/2,n}}{%
J_{ma+1/2}^{2}(\lambda _{ma-1/2,n})} &=&\frac{1}{2}\int_{0}^{\infty
}dx\,f(x)-\frac{1}{2\pi }\int_{0}^{\infty }dx\,\frac{K_{ma-1/2}(x)}{%
I_{ma-1/2}(x)}  \notag \\
&&\times \left[ e^{\left( 1/2-ma\right) \pi i}f(ix)+e^{\left( ma-1/2\right)
\pi i}f(-ix)\right] .  \label{SummAP}
\end{eqnarray}%
The part of the VEV corresponding to the first term in the right-hand side
of (\ref{SummAP}) gives the VEV in the geometry without the brane. For the
brane-induced contribution coming from the second integral in (\ref{SummAP})
we get%
\begin{equation}
\langle T_{\mu }^{\rho }\rangle _{\mathrm{b}}^{\mathrm{(f)}}=-\frac{%
2^{-D}\delta _{\mu }^{\rho }N}{\pi ^{D/2}\Gamma (D/2)a^{D+1}}%
\int_{0}^{\infty }dx\,x^{D+1}\frac{K_{ma-1/2}(xz_{0}/z)}{I_{ma-1/2}(xz_{0}/z)%
}F_{\mathrm{(L)}}^{(\mu )}(x),  \label{EMTb2}
\end{equation}%
where the functions in the integrand are defined as
\begin{eqnarray}
F_{\mathrm{(L)}}^{(0)}(x) &=&\frac{1}{D}\left[
I_{ma-1/2}^{2}(x)-I_{ma+1/2}^{2}(x)\right] ,  \notag \\
F_{\mathrm{(L)}}^{(D)}(x) &=&I_{ma+1/2}^{2}(x)-I_{ma-1/2}^{2}(x)+\frac{2ma}{x%
}I_{ma+1/2}(x)I_{ma-1/2}(x).  \label{S2nu}
\end{eqnarray}%
The energy density and the parallel stresses corresponding to (\ref{EMTb2})
are negative, $\langle T_{\mu }^{\mu }\rangle _{\mathrm{b}}^{\mathrm{(f)}}<0$
(no summation over $\mu $), $\mu =0,1,\ldots ,D-1$, whereas the normal
stress is positive. Comparing with the results for the R-region, we see that
the energy densities in the R- and L-regions have the same sign and the
normal stresses have opposite signs.

For a massless field one obtains $\langle T_{D}^{D}\rangle _{\mathrm{b}}^{%
\mathrm{(f)}}=-D\langle T_{0}^{0}\rangle _{\mathrm{b}}^{\mathrm{(f)}}$ and
the expression for the energy density is simplified to
\begin{equation}
\langle T_{0}^{0}\rangle _{\mathrm{b}}^{\mathrm{(f)}}=-\left( \frac{z}{az_{0}%
}\right) ^{D+1}\frac{N\Gamma ((D+1)/2)}{\left( 4\pi \right) ^{(D+1)/2}}%
(1-2^{-D})\zeta (D+1),  \label{T00Lm0}
\end{equation}%
where $\zeta (x)$ is the Riemann zeta function. In this case the
brane-induced part is traceless. The massless fermionic field is conformally
invariant and (\ref{T00Lm0}) corresponds to the conformal relation (\ref%
{ConfRel}). In the Minkowskian counterpart, $z_{0}$ is the separation
between two parallel planar boundaries. For a massive field and in the
Minkowskian limit $a\rightarrow 0$ we obtain the expression (\ref{EMTRD})
for the energy density and parallel stresses, with the replacement $%
y-y_{0}\rightarrow y_{0}-y$, and the normal stress vanishes.

The VEV (\ref{EMTb2}) depends on $z$ and $z_{0}$ in the form of the ration $%
z/z_{0}$. Let us consider its behavior in the asymptotic regions. The limit $%
z/z_{0}\rightarrow 1$ corresponds to the points on the brane. The
brane-induced contribution (\ref{EMTb2}) diverges on the brane. The leading
terms in the asymptotic expansions over the distance from the brane are
obtained from the corresponding expressions (\ref{EMTnear}) for the R-region
by the replacement $y-y_{0}\rightarrow y_{0}-y$ and by an additional change
of the sign for the normal stress. For $z/z_{0}\ll 1$ we use the small
argument asymptotics for the functions $F_{\mathrm{(L)}}^{(\mu )}(x)$. To
the leading order this gives%
\begin{equation*}
\langle T_{0}^{0}\rangle _{\mathrm{b}}^{\mathrm{(f)}}\approx -\frac{%
2^{-D-2ma}\pi ^{-D/2}N(z/z_{0})^{D+2ma+1}}{\Gamma (D/2+1)\Gamma ^{2}\left(
ma+1/2\right) a^{D+1}}\int_{0}^{\infty }dx\,x^{D+2ma}\frac{K_{ma-1/2}(x)}{%
I_{ma-1/2}(x)},
\end{equation*}%
for the energy density and
\begin{equation*}
\langle T_{D}^{D}\rangle _{\mathrm{b}}^{\mathrm{(f)}}\approx -\frac{D}{2ma+1}%
\langle T_{0}^{0}\rangle _{\mathrm{b}}^{\mathrm{(f)}},
\end{equation*}%
for the normal stress. In particular, the brane-induced VEV tends to zero on
the AdS boundary as $z^{D+2ma+1}$.

\begin{figure}[tbph]
\begin{center}
\begin{tabular}{cc}
\epsfig{figure=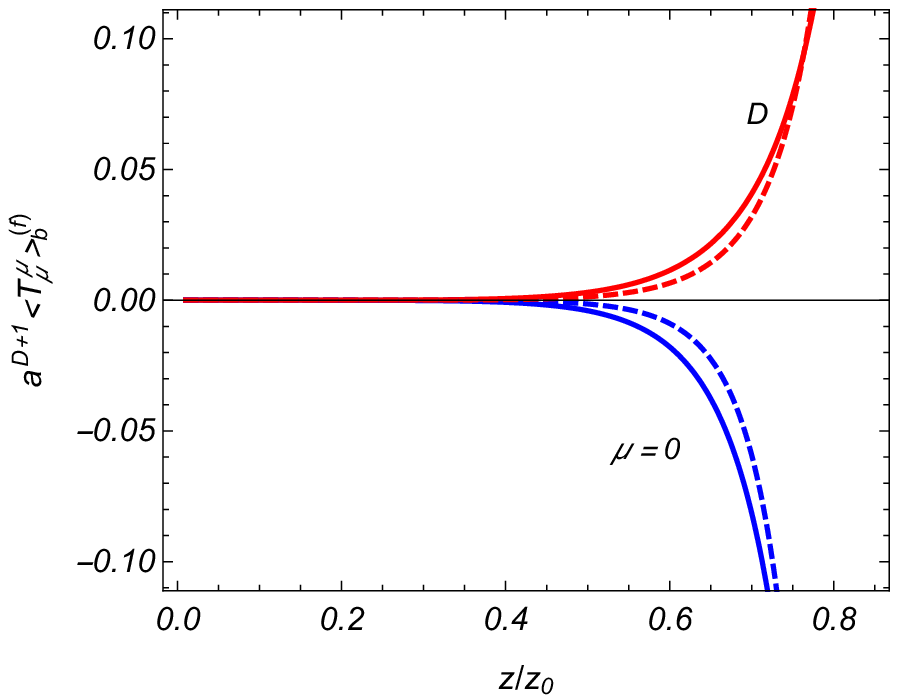,width=7.cm,height=5.5cm} & \quad %
\epsfig{figure=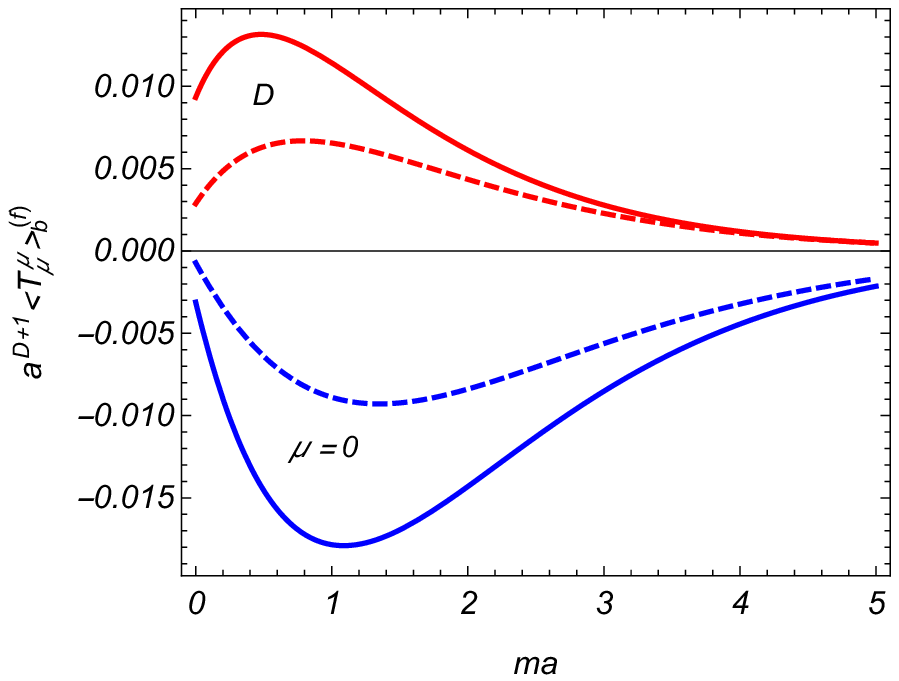,width=7.cm,height=5.5cm}%
\end{tabular}%
\end{center}
\caption{The same as in figure \protect\ref{fig1} for the L-region. The left
and right panels are plotted for $ma=1$ and $z/z_{0}=0.6$, respectively.}
\label{fig5}
\end{figure}

The brane-induced contributions for the energy density and the normal stress
in the L-region, $\langle T_{\mu }^{\mu }\rangle _{\mathrm{b}}^{\mathrm{(f)}}
$ (no summation over $\mu $), $\mu =0,D$, for $D=3$ (full curves) and $D=4$
(dashed curves) Dirac fields are presented in figure \ref{fig5} as functions
of $z/z_{0}$ and of $ma$. The left and right panels are plotted for $ma=1$
and $z/z_{0}=0.6$.

\subsubsection{Scalar field}

For a scalar field the mode-sum formula for the VEV\ of the energy-momentum
tensor contains series over the zeros of the function $\bar{J}_{\nu
}(x)=B_{0}xJ_{\nu }^{\prime }(x)+A_{0}J_{\nu }(x)$. The corresponding
summation formula can be found in \cite{Saha87,Saha08}. The brane-induced
contribution for a scalar field is presented as%
\begin{equation}
\langle T_{\mu }^{\rho }\rangle _{\mathrm{b}}^{\mathrm{(s)}}=-\frac{%
2^{-D}\delta _{\mu }^{\rho }}{\pi ^{D/2}\Gamma (D/2)a^{D+1}}\int_{0}^{\infty
}dx\,x^{D+1}\frac{\bar{K}_{\nu }(xz_{0}/z)}{\bar{I}_{\nu }(xz_{0}/z)}S^{(\mu
)}\left[ I_{\nu }(x)\right] ,  \label{TmuSL}
\end{equation}%
where the functions $S^{(\mu )}[g(x)]$ are given by (\ref{SDnew}). For a
conformally coupled massless field one has $\nu =1/2$ and $%
S^{(0)}[I_{1/2}(x)]=2/(\pi Dx)$, $S^{(D)}[I_{1/2}(x)]=-2/(\pi x)$. For the
energy density we get
\begin{equation}
\langle T_{0}^{0}\rangle _{\mathrm{b}}^{\mathrm{(s)}}=-\frac{%
(z/a)^{D+1}z_{0}^{-D-1}}{2^{2D+1}\pi ^{D/2}\Gamma (D/2+1)}\int_{0}^{\infty
}dx\frac{x^{D}}{\frac{2a/\beta +1-D-x}{2a/\beta +1-D+x}e^{x}-1},
\label{T00SM}
\end{equation}%
and for the normal stress $\langle T_{D}^{D}\rangle _{\mathrm{b}}^{\mathrm{%
(s)}}=-D\langle T_{0}^{0}\rangle _{\mathrm{b}}^{\mathrm{(s)}}$. Note that
one has the conformal relation $\varphi (x)=(z/a)^{(D-1)/2}\varphi _{\mathrm{%
M}}(x)$ between the fields in the AdS and Minkowski bulk. From here it
follows that the conformal image of the boundary condition (\ref{Rob}) is
the condition $(\beta _{\mathrm{M}}n_{\mathrm{M}}^{\mu }\nabla _{\mu
}+1)\varphi _{\mathrm{M}}(x)=0$, $z=z_{0}$, with the relation between the
Robin coefficients%
\begin{equation}
\frac{z_{0}}{\beta _{\mathrm{M}}}=\frac{a}{\beta }-\frac{D-1}{2}.
\label{RelRob}
\end{equation}%
By taking into account (\ref{RelRob}), we can see that (\ref{T00SM})
corresponds to the conformal relation (\ref{ConfRel}) with the Minkowskian
problem of two boundaries with the Dirichlet boundary condition on the plate
$z=0$ (conformal image of the AdS boundary) and the Robin condition with the
coefficient $\beta _{\mathrm{M}}$ on the plate at $z=z_{0}$. In the
Minkowskian limit the result (\ref{TmuM}) is obtained for the energy density
and parallel stresses and the normal stress becomes zero.

Let us consider the behavior of the brane-induced VEV (\ref{TmuSL}) near the
brane and near the AdS boundary. Near the brane one has $1-z/z_{0}\ll 1$ and
the contribution of large $x$ dominates in the integral of (\ref{TmuSL}).
For non-conformally coupled fields the leading term in the expansion of the
energy density is given by (\ref{T00nearS}) with the replacement $%
y-y_{0}\rightarrow y_{0}-y$ and the relation between the energy density and
the normal stress remain the same. In particular, we see that near the brane
the energy density has the same sign in the R- and L-regions, whereas the
normal stresses have opposite signs. For points near the AdS boundary one
has $z/z_{0}\ll 1$ and the leading term in the asymptotic expansion is given
by%
\begin{equation}
\langle T_{0}^{0}\rangle _{\mathrm{b}}^{\mathrm{(s)}}\approx \frac{%
(z/z_{0})^{D+2\nu }}{\pi ^{D/2}\Gamma (D/2)}\frac{\left( D+2\nu +1\right)
\xi _{1}+1}{2^{D+2\nu }\Gamma (\nu )\Gamma (\nu +1)a^{D+1}}\int_{0}^{\infty
}dxx^{D+2\nu -1}\frac{\bar{K}_{\nu }(x)}{\bar{I}_{\nu }(x)}.
\label{TmunearB}
\end{equation}%
In the same order, the normal stress is found from $\langle T_{D}^{D}\rangle
_{\mathrm{b}}^{\mathrm{(s)}}\approx -D\langle T_{0}^{0}\rangle _{\mathrm{b}%
}^{\mathrm{(s)}}/(2\nu )$. Hence, the brane-induced VEVs vanish on the AdS
boundary like $z^{D+2\nu }$. Note that the factor $\left( D+2\nu +1\right)
\xi _{1}+1$ is negative for both minimally and conformally coupled field and
for the Dirichlet boundary condition the energy density is negative near the
AdS boundary. In general, depending on the Robin coefficient, the energy
density can be either negative or positive.

\begin{figure}[tbph]
\begin{center}
\begin{tabular}{cc}
\epsfig{figure=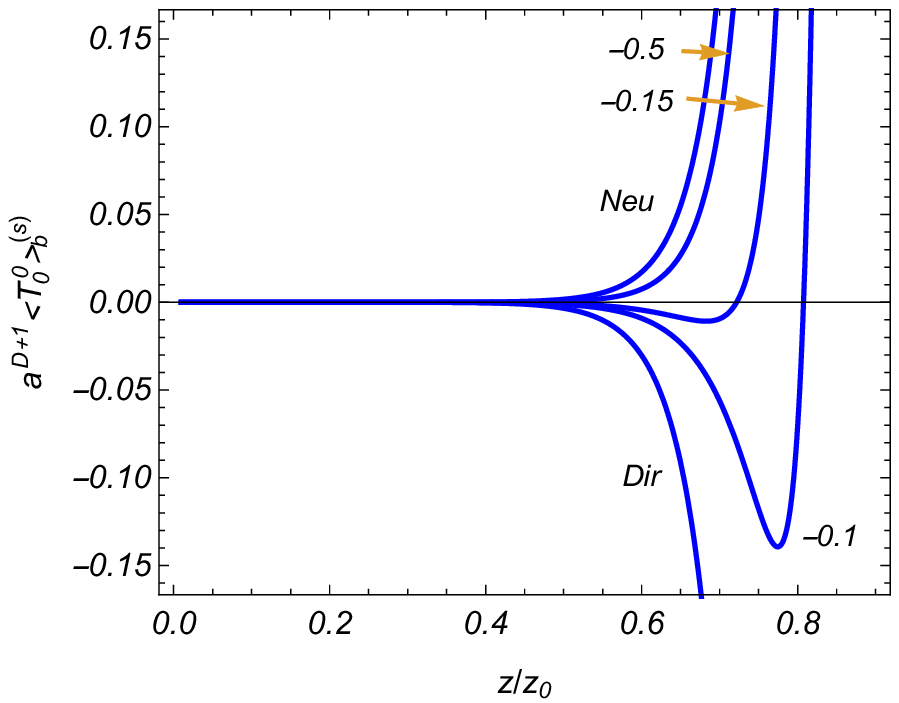,width=7.cm,height=5.5cm} & \quad %
\epsfig{figure=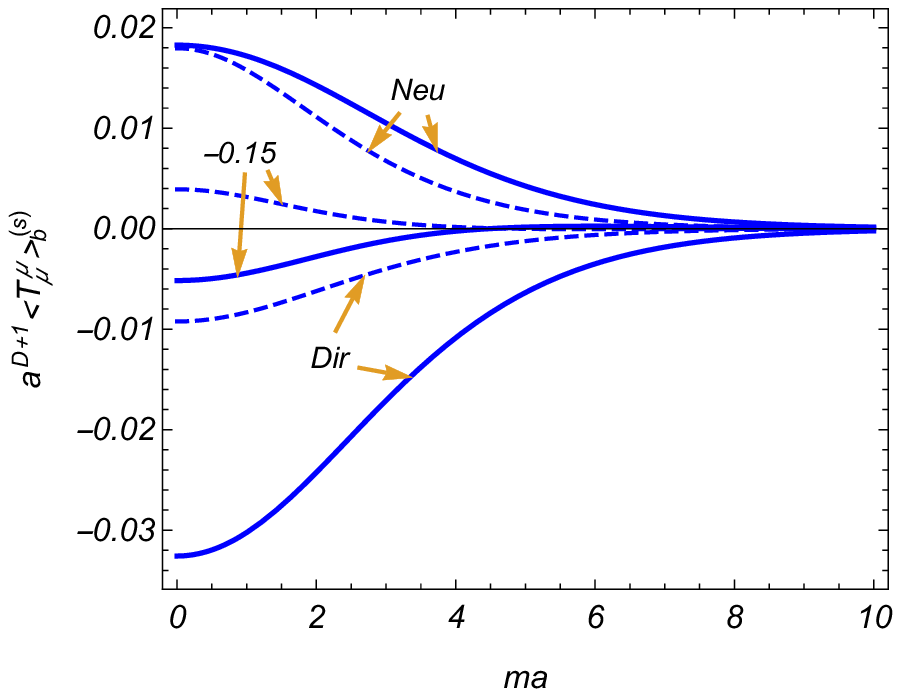,width=7.cm,height=5.5cm}%
\end{tabular}%
\end{center}
\caption{The same as in figure \protect\ref{fig2} for a minimally coupled
scalar field in the L-region. The left and right panels are plotted for $ma=1
$ and $z/z_{0}=0.6$, respectively.}
\label{fig6}
\end{figure}

The brane-induced energy density for $D=4$ minimally coupled scalar field is
presented on the left panel of figure \ref{fig6} as a function of $z/z_{0}$.
The graphs are plotted for $ma=1$, for the Dirichlet and Neumann boundary
conditions (the labels Dir and Neu near the curves), and for the Robin
boundary conditions with $\beta /a=-0.5,-0.15,-0.1$. For the latter cases
the energy density $\langle T_{0}^{0}\rangle _{\mathrm{b}}^{\mathrm{(s)}}$
increases with increasing $|\beta /a|$. On the right panel we display the
energy density and the normal stress as functions of the mass for the
Dirichlet and Neumann boundary conditions, and for the Robin condition with $%
\beta /a=-0.15$. The graphs are plotted for $z/z_{0}=0.6$. For the Robin
boundary condition the brane-induced VEVs may change the sign as functions
of the mass.

It is of interest to compare the boundary-induced VEVs in AdS and dS
spacetimes. The scalar Casimir densities in background of dS spacetime have
been investigated in \cite{Saha08dS} and \cite{Eliz10dS} for a single and
two parallel Robin boundaries, respectively. The corresponding line element
is taken in planar coordinates,
\begin{equation}
ds^{2}=dt^{2}-e^{2t/a}\sum_{i=1}^{D}(dz^{i})^{2},  \label{dsdS}
\end{equation}%
with $-\infty <t,z^{i}<+\infty $, and it is assumed that the field is
prepared in the Bunch-Davies vacuum state. In the geometry of a single
boundary at $z^{D}=0$, the boundary-induced VEV of the energy-momentum
tensor is expressed in terms of integrals that contain the products of the
modified Bessel functions with the order $\nu _{\mathrm{dS}}=\left[
D^{2}/4-D(D+1)\xi -m^{2}a^{2}\right] ^{1/2}$ (compare with (\ref{nu})). The
VEV, in addition to the diagonal components, has also nonzero off-diagonal
component $\langle T_{0}^{D}\rangle _{\mathrm{b}}^{\mathrm{(s)}}$ that
describes the energy flux along the direction normal to the boundary.
Depending on the boundary condition, the flux can be either positive or
negative. Another qualitatively new effect of the gravity, compared with the
corresponding problem in the Minkowski bulk, is the appearance of the
nonzero normal stress $\langle T_{D}^{D}\rangle _{\mathrm{b}}^{\mathrm{(s)}}$%
. All the components $\langle T_{\mu }^{\rho }\rangle _{\mathrm{b}}^{\mathrm{%
(s)}}$ depend on the spacetime coordinates $t$ and $z^{D}$ in the form of
the combination $|z^{D}|e^{t/a}$. The latter is the proper distance from the
boundary measured in units of the curvature radius $a$. Similar to the case
of the AdS bulk, for small values of that combination the influence of the
gravitational field on the energy density and stresses parallel to the
boundary is weak. At large proper distances from the boundary, $%
|z^{D}|e^{t/a}\gg a$, the decay of the boundary-induced VEV is qualitatively
different for real and purely imaginary values of $\nu _{\mathrm{dS}}$. For $%
\nu _{\mathrm{dS}}>0$ the VEV tends to zero monotonically, like $%
(|z^{D}|e^{t/a})^{2\nu _{\mathrm{dS}}-D-\chi }$, where $\chi =0$ for the
diagonal components and $\chi =1$ for the component $\langle
T_{0}^{D}\rangle _{\mathrm{b}}^{\mathrm{(s)}}$. For imaginary $\nu _{\mathrm{%
dS}}$ the behavior of the boundary-induced VEV at large proper distances is
damping oscillatory, like $(|z^{D}|e^{t/a})^{-D-\chi }\sin \left[ 2|\nu _{%
\mathrm{dS}}|\ln (|z^{D}|e^{t/a})+\phi \right] $, with $\phi $ being a
constant phase. Recall that for the AdS bulk the decay of the
boundary-induced VEV of the energy-momentum tensor, as a function of the
proper distance $|y-y_{0}|$, was exponential at large distances (see (\ref%
{T00farS}), (\ref{TmunearB})).

\subsubsection{Electromagnetic field}

For the electromagnetic field the eigenmodes of the quantum number $\lambda $
are roots of the equation (\ref{JnuE}), where $\nu $ is given by (\ref{nuV})
for the boundary conditions (\ref{BC}) and (\ref{BC2}). The corresponding
summation formula for the series in the mode-sum of the energy-momentum
tensor is obtained from (\ref{SummAP}) by the replacement $ma-1/2\rightarrow
\nu $. The brane-induced contribution is expressed as%
\begin{equation}
\langle T_{\mu }^{\rho }\rangle _{\mathrm{b}}^{\mathrm{(v)}}=-\frac{\delta
_{\nu }\delta _{\mu }^{\rho }\left( D-1\right) }{2^{D}\pi ^{D/2}\Gamma
(D/2)a^{D+1}}\int_{0}^{\infty }dx\,x^{D+1}\frac{K_{\nu }(xz_{0}/z)}{I_{\nu
}(xz_{0}/z)}V_{\mathrm{(L)}}^{(\mu )}(x),  \label{TmuVL}
\end{equation}%
where we have defined the functions
\begin{eqnarray}
V_{\mathrm{(L)}}^{(0)}(x) &=&\left( 1-\frac{4}{D}\right)
I_{D/2-1}^{2}(x)+\left( 1-\frac{2}{D}\right) I_{D/2-2}^{2}(x),  \notag \\
V_{\mathrm{(L)}}^{(D)}(x) &=&I_{D/2-1}^{2}(x)-I_{D/2-2}^{2}(x).  \label{VDL}
\end{eqnarray}%
For $D\geq 4$ the energy density is negative for the boundary condition (\ref%
{BC}) and positive for the condition (\ref{BC2}). For the normal stress one
has $\langle T_{D}^{D}\rangle _{\mathrm{b}}^{\mathrm{(v)}}>0$ in the case of
the condition (\ref{BC}) and $\langle T_{D}^{D}\rangle _{\mathrm{b}}^{%
\mathrm{(v)}}<0$ for (\ref{BC2}). In the Minkowskian limit $a\rightarrow
\infty $, with fixed $y,y_{0}$, we obtain the result (\ref{TmuVM}).

Unlike the R-region, in the L-region the brane-induced contribution is
different from zero for $D=3$. For the boundary condition (\ref{BC}) one gets%
\begin{equation}
\langle T_{\mu }^{\rho }\rangle _{\mathrm{b}}^{\mathrm{(v)}}=-\frac{\pi ^{2}%
}{720}\left( \frac{z}{az_{0}}\right) ^{4}\mathrm{diag}(1,1,1,-3).
\label{TmuVD3}
\end{equation}%
This determines the Casimir force per unit surface of the brane equal to $%
-\pi ^{2}a^{-4}/240$. The latter is attractive with respect to the AdS
boundary and does not depend on the location of the brane. In the case of
the boundary condition (\ref{BC2}) we obtain the expression%
\begin{equation}
\langle T_{\mu }^{\rho }\rangle _{\mathrm{b}}^{\mathrm{(v)}}=\frac{7\pi ^{2}%
}{5760}\left( \frac{z}{az_{0}}\right) ^{4}\mathrm{diag}(1,1,1,-3).
\label{TmuVD3b}
\end{equation}%
The corresponding force per unit surface of the brane is given by $7\pi
^{2}\alpha ^{-4}/1920$ and is repulsive with respect to the AdS boundary.

The near-brane asymptotics in the L-region are given by (\ref{T00near}) with
the replacement $y-y_{0}\rightarrow y_{0}-y$ in the expression for the
energy density. Near the AdS boundary one has $z/z_{0}\ll 1$ and by using
the expression of the modified Bessel function for small arguments, in the
leading order we get%
\begin{eqnarray}
\langle T_{0}^{0}\rangle _{\mathrm{b}}^{\mathrm{(v)}} &\approx &-\frac{%
8\delta _{\nu }\left( D-1\right) \left( z/2z_{0}\right) ^{2D-2}}{\pi
^{D/2}D\Gamma ^{3}(D/2-1)a^{D+1}}\int_{0}^{\infty }dx\,x^{2D-3}\frac{K_{\nu
}(x)}{I_{\nu }(x)},  \notag \\
\langle T_{D}^{D}\rangle _{\mathrm{b}}^{\mathrm{(v)}} &\approx &-\frac{D}{D-2%
}\langle T_{0}^{0}\rangle _{\mathrm{b}}^{\mathrm{(v)}}.  \label{TDDVb}
\end{eqnarray}%
In spatial dimension $D=3$ the integral is equal to $\pi ^{5}/240$ for the
boundary condition (\ref{BC}) and $7\pi ^{5}/1920$ for (\ref{BC2}). In this
special case the asymptotics (\ref{TDDVb}) coincide with the exact results (%
\ref{TmuVD3}) and (\ref{TmuVD3b}).

\begin{figure}[tbph]
\begin{center}
\begin{tabular}{cc}
\epsfig{figure=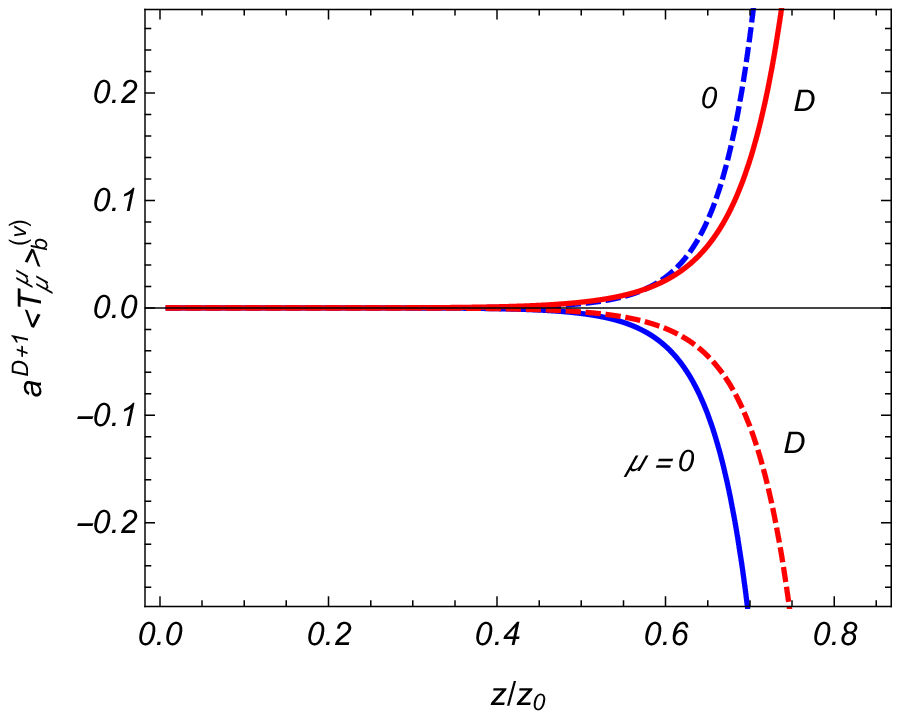,width=7.cm,height=5.5cm} & \quad %
\epsfig{figure=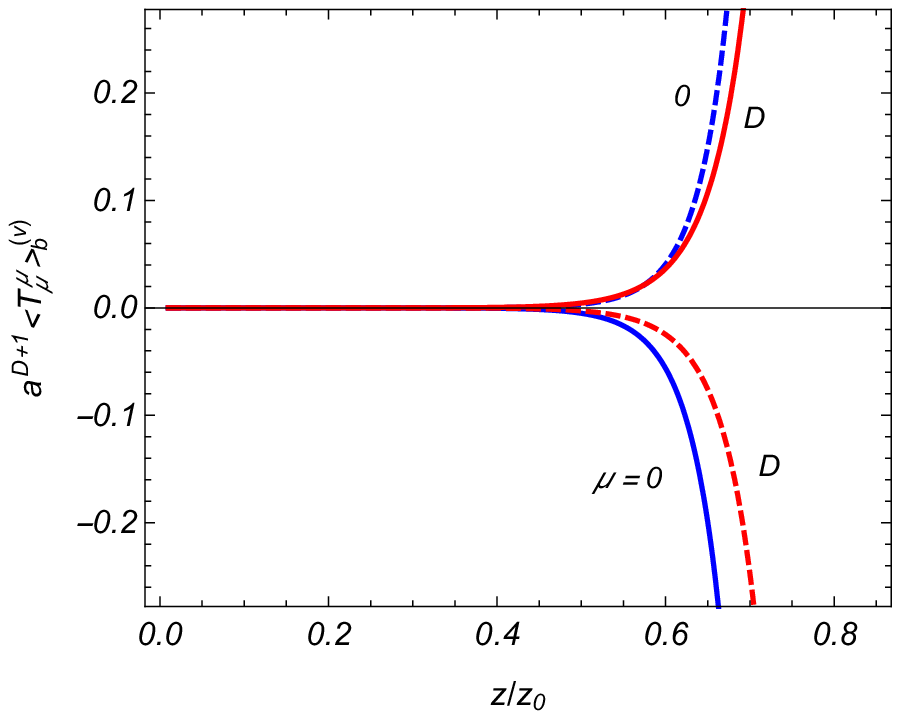,width=7.cm,height=5.5cm}%
\end{tabular}%
\end{center}
\caption{The brane-induced VEVs of the energy density and the normal stress
in the L-region as functions of $z/z_{0}$ for $D=4$ (left panel) and $D=5$
(right panel) electromagnetic fields. The full and dashed curves correspond
to the conditions (\protect\ref{BC}) and (\protect\ref{BC2}).}
\label{fig7}
\end{figure}

For $D=4$ and $D=5$ electromagnetic fields, the brane contributions to the
vacuum energy density and the normal stress are plotted on the left and
right panels of figure \ref{fig7} (the values of the index $\mu $ for $%
\langle T_{\mu }^{\mu }\rangle _{\mathrm{b}}^{\mathrm{(v)}}$ (no summation
over $\mu $) are displayed near the curves). The full and dashed curves
correspond to the boundary conditions (\ref{BC}) and (\ref{BC2}),
respectively.

The electromagnetic Casimir densities for the boundary condition (\ref{BC})
on a single and two parallel plates in dS spacetime with the line element (%
\ref{dsdS}) have been investigated in \cite{Saha14dS,Saha15dS}. It was
assumed that the field is prepared in the Bunch-Davies vacuum. For $D=3$,
the electromagnetic field is conformally invariant and the plate-induced
contribution in the VEV\ of the energy-momentum tensor is conformally
related to that for the Minkowski bulk. In particular, it vanishes in the
geometry of a single plate. In spatial dimensions $D\geq 4$, the components
of the plate-induced VEV are expressed in terms of the hypergeometric
function. For these dimensions, similar to the case of a scalar field, the
VEV of the energy-momentum tensor has nonzero off-diagonal component $%
\langle T_{0}^{D}\rangle _{\mathrm{b}}^{\mathrm{(v)}}$. For a single plate,
the corresponding energy flux is directed from the plate. At large proper
distances from the plate, located at $z^{D}=0$, the vacuum stresses are
isotropic and for $D>4$ the diagonal components decay as $\left(
|z^{D}|e^{t/a}\right) ^{-4}$ and the off-diagonal component behaves like $%
(|z^{D}|e^{t/a})^{-5}$. For $D=4$ the asymptotics for $\langle
T_{0}^{0}\rangle _{b}$ and $\langle T_{0}^{D}\rangle _{b}$ remain the same
and the stresses behave as $(|z^{D}|e^{t/a})^{-6}\ln \left(
|z^{D}|e^{t/a}\right) $. In dS spacetime we have a power-law decay of the
boundary-induced VEV as a function of the proper distance from the plate. In
the AdS bulk the decay is exponential.

\section{VEV\ of the surface energy-momentum tensor for a scalar field}

\label{sec:Surf}

In the discussion above we have considered the VEV of the bulk
energy-momentum tensor. In manifolds with boundaries, in addition to the
latter a surface energy-momentum tensor may present which is localized on
the boundaries. In the general case of bulk and boundary geometries, the
expression of the surface energy-momentum tensor for a scalar field with
general curvature coupling has been obtained in \cite{Saha04Surf} by using
the standard variational procedure. The VEV of the surface energy-momentum
tensor for branes parallel to the AdS boundary is investigated in \cite%
{Saha04b} by using the generalized zeta function technique.

For a given field, the expression for the surface energy-momentum tensor $%
T_{\mu \rho }^{\mathrm{(surf)}}$, in addition to the bulk action, depends on
the surface action. In \cite{Saha04Surf}, for a spacetime region $M$ with
boundary $\partial M$ the surface action for a scalar field is taken in the
form%
\begin{equation}
S_{s}=-\epsilon \int_{\partial M}d^{D}x\sqrt{|h|}\left( \xi K+m_{s}\right)
\varphi ^{2},  \label{Sact}
\end{equation}%
where $\epsilon =1$ for spacelike and $\epsilon -1$ for timelike elements of
the boundary, $h$ is the determinant of the induced metric $h_{\mu \rho
}=g_{\mu \rho }-\epsilon n_{\mu }n_{\rho }$, with $n_{\mu }$ being the
inward pointing unit normal to $\partial M$, $n_{\mu }n^{\mu }=\epsilon $.
In (\ref{Sact}), $K=g^{\mu \rho }K_{\mu \rho }$ is the trace of the
extrinsic curvature tensor $K_{\mu \rho }=h_{\mu }^{\sigma }h_{\rho }^{\tau
}\nabla _{\sigma }n_{\tau }$ of the boundary and $m_{s}$ is a parameter. $%
\partial M$ consists of the initial and final spacelike hypersurfaces and a
timelike smooth boundary $\partial M_{s}$. The variation of the total action
with respect to the field $\varphi (x)$ leads to the standard field equation
(\ref{fieldeq}) in the bulk and to the boundary condition
\begin{equation}
\left( 2\xi K+2m_{s}+n^{\mu }\nabla _{\mu }\right) \varphi (x)=0,\;x\in
\partial M_{s}.  \label{BCgen}
\end{equation}%
The variation of the action with respect to the metric tensor gives the
metric energy-momentum tensor. In addition to the bulk part, the latter
contains a contribution $T_{\mu \rho }^{\mathrm{(surf)}}$ located on the
boundary $\partial M_{s}$: $T_{\mu \rho }^{\mathrm{(surf)}}=\tau _{\mu \rho
}\delta (x;\partial M_{s})$, where $\delta (x;\partial M_{s})$ is the
'one-sided' $\delta $-function. By using the boundary condition (\ref{BCgen}%
) the expression for $\tau _{\mu \rho }$ is presented in the form \cite%
{Saha04Surf}%
\begin{equation}
\tau _{\mu \rho }=\xi \varphi ^{2}K_{\mu \rho }-\left( 2\xi -\frac{1}{2}%
\right) h_{\mu \rho }\varphi n^{\sigma }\nabla _{\sigma }\varphi .
\label{tausurf}
\end{equation}%
Note that the boundary condition (\ref{BCgen}) is of the Robin type. By
using the boundary condition, one can exclude the derivative term for the
field in (\ref{tausurf}).

In the geometry (\ref{metric2}) with a single brane at $z=z_{0}$, the
extrinsic curvature tensor for the R- and L-regions (J=R,L) has the form $%
K_{\mu \rho }^{\mathrm{(J)}}=-\delta _{\mathrm{(J)}}g_{\mu \rho }/a$ for $%
\mu ,\rho =0,1,\ldots ,D-1$, and $K_{DD}^{\mathrm{(J)}}=0$. The boundary
condition (\ref{BCgen}) is reduced to (\ref{Rob}) with $1/\beta
=2m_{s}-2\delta _{\mathrm{(J)}}D\xi /a$. The VEV of the surface
energy-momentum tensor, $\left\langle 0\right\vert \tau _{\mu \rho
}\left\vert 0\right\rangle \equiv \left\langle \tau _{\mu \rho
}\right\rangle $, is evaluated by using the mode-sum formula $\left\langle
\tau _{\mu \rho }\right\rangle =\sum_{\sigma }\sum_{s=\pm }\tau _{\mu \rho
}\{\varphi _{\sigma }^{(s)}(x),\varphi _{\sigma }^{(s)\ast }(x)\}/2$ with
the mode functions given by (\ref{ModesR}) and (\ref{ModesL}) for the R- and
L-regions, respectively. The VEV has the form $\left\langle \tau _{\mu
}^{\rho }\right\rangle =\mathrm{const}\cdot \delta _{\mu }^{\rho }$, $\mu
,\rho =0,1,\ldots ,D-1$, $\left\langle \tau _{\mu }^{D}\right\rangle =0$
and, from the point of view of an observer living on the brane it
corresponds to a gravitational source of the cosmological constant type. An
essential difference compared with the bulk energy-momentum tensor is that
the subtraction of the part corresponding to the geometry without a brane is
not sufficient and an additional renormalization is required. The latter is
reduced to the renormalization of the VEV\ for the field squared on the
brane. In \cite{Saha04b} the generalized zeta function technique has been
used.

The VEV of the surface energy-momentum tensor for the region J=R,L, $%
\left\langle \tau _{\mu \rho }\right\rangle ^{\mathrm{(J)}}$, is expressed
in terms of the VEV\ of the field squared on the brane, $\left\langle
\varphi ^{2}\right\rangle _{z=z_{0}}^{\mathrm{(J)}}$, as%
\begin{equation}
\left\langle \tau _{\mu }^{\rho }\right\rangle ^{\mathrm{(J)}}=\delta _{\mu
}^{\rho }\delta _{\mathrm{(J)}}\left[ (2\xi -1/2)/\beta -\xi /a\right]
\left\langle \varphi ^{2}\right\rangle _{z=z_{0}}^{\mathrm{(J)}},
\label{tauj}
\end{equation}%
The VEV $\left\langle \varphi ^{2}\right\rangle _{z=z_{0}}^{\mathrm{(J)}}$
is obtained by the analytic continuation of the function (the details can be
found in \cite{Saha04b})%
\begin{equation}
F_{\mathrm{(J)}}(s)=-\frac{\delta _{\mathrm{(J)}}(\sqrt{4\pi }a)^{1-D}\beta
(\mu z_{0})^{-s-1}}{\Gamma \left( -s/2\right) \Gamma \left( (D+1+s)/2\right)
}\int_{0}^{\infty }dx\,x^{D+s}U_{\mathrm{(J)}\nu }(x),  \label{Fab}
\end{equation}%
to the physical point $s=-1$. Here, the parameter $\mu $ is the
renormalization scale, $\nu $ is defined by (\ref{nu}) and%
\begin{equation}
U_{\mathrm{(R)}\nu }(x)=\frac{K_{\nu }(x)}{\bar{K}_{\nu }(x)},\quad U_{%
\mathrm{(L)}\nu }(x)=\frac{I_{\nu }(x)}{\bar{I}_{\nu }(x)}.  \label{URL}
\end{equation}%
The representation (\ref{Fab}) is valid in the slice $-(D+1)<\mathrm{Re}%
\,s<-D$ of the complex $s$-plane. In the first step of the analytic
continuation the integral in (\ref{Fab}) is presented in the form of the sum
of the integrals over the regions $[0,1]$ and $[1,\infty )$. In the first
integral the substitution $s=-1$ can be made directly. In the second
integral we subtract and add to the function $U_{\mathrm{(J)}\nu }(x)$ in
the integrand the $N$ leading terms of the corresponding asymptotic
expansion for large values of $x$ and integrate the asymptotic part. For $%
\beta \neq 0$ the asymptotic expansion has the form $\beta U_{\mathrm{(J)}%
\nu }(x)\sim \sum_{l=0}^{\infty }w_{l}(\nu )(-\delta _{\mathrm{(J)}%
}x)^{-l-1} $, where the coefficients $w_{l}(\nu )$ are found from those for
the expansions of the modified Bessel functions. The function $F_{\mathrm{(J)%
}}(s)$ has a simple pole at $s=-1$ and the leading term in the Laurent
expansion is given by
\begin{equation}
\frac{-2(-\delta _{\mathrm{(J)}}a)^{1-D}w_{D-1}(\nu )}{(4\pi )^{D/2}\Gamma
\left( D/2\right) (s+1)}.  \label{Pole}
\end{equation}%
In this way, the VEVs $\left\langle \varphi ^{2}\right\rangle _{z=z_{0}}^{%
\mathrm{(J)}}$ for J=R and J=L are decomposed into the pole and finite
contributions. The pole terms can be absorbed by adding to the brane action
the respective counterterms. The expressions for the finite parts in
separate regions will not be given here and can be found in \cite{Saha04b}.
We will consider the total energy density.

Combining the results for the R- and L-regions, one obtains the total
surface energy density $\left\langle \tau _{0}^{0}\right\rangle
=\left\langle \tau _{0}^{0}\right\rangle ^{\mathrm{(R)}}+\left\langle \tau
_{0}^{0}\right\rangle ^{\mathrm{(L)}}$. Comparing the pole parts (\ref{Pole}%
) for $\left\langle \varphi ^{2}\right\rangle _{z=z_{0}}^{\mathrm{(J)}}$ in
those regions, we can see that in odd spatial dimensions the pole parts in
the energy density cancel out and the finite part does not depend on the
renormalization scale $\mu $. Taking $N=D-1$ for the number of the terms
taken in the asymptotic expansions of the function $U_{\mathrm{(J)}\nu }(x)$%
, for the total surface energy density in odd dimensions $D$ one gets the
formula%
\begin{eqnarray}
\left\langle \tau _{0}^{0}\right\rangle &=&\frac{2\xi \beta /a+1-4\xi }{%
(4\pi )^{D/2}\Gamma \left( D/2\right) a^{D}}\left[ \int_{0}^{1}dx\,x^{D-1}%
\sum_{\mathrm{J=R,L}}U_{\mathrm{(J)}\nu }(x)-\frac{2}{\beta }%
\sum_{l=0}^{(D-3)/2}\frac{w_{2l+1}(\nu )}{D-2l-2}\right.  \notag \\
&&\left. +\int_{1}^{\infty }dx\,x^{D-1}\left( \sum_{\mathrm{J=R,L}}U_{%
\mathrm{(J)}\nu }(x)-\frac{2}{\beta }\sum_{l=0}^{(D-3)/2}\frac{w_{2l+1}(\nu )%
}{x^{2l+2}}\right) \right] .  \label{tau00}
\end{eqnarray}
Note that this quantity does not depend on the location of the brane.
Depending on the value of the Robin coefficient $\beta $, the surface energy
density (\ref{tau00}) can be either positive or negative (see the graphs in
\cite{Saha04b} for minimally and conformally coupled scalar fields).

In the geometry of two branes, the VEV of the surface energy-momentum tensor
on a given brane is decomposed into two parts. The first one corresponds to
the VEV in the problem where the second brane is absent and the
corresponding evaluation procedure has been described in this section. The
second part is induced by the presence of the second brane and it requires
no additional renormalization. As it has been discussed in \cite{Saha04b},
in the Randall-Sundrum model the surface energy density induced on the
visible brane by the presence of the hidden brane gives rise to naturally
suppressed cosmological constant. The surface energy density in models with
additional compact dimensions is discussed in \cite{Saha06,Saha18} for
neutral and charged scalar fields. In the latter case, the value of the
induced cosmological constant on the brane is additionally controlled by
tuning the magnetic flux enclosed by compact dimensions.

\section{Geometry with a brane perpendicular to the AdS boundary}

\label{sec:PerpBr}

In a number of recent developments of the AdS/CFT correspondence, branes
intersecting the AdS boundary are considered. They include the extensions of
the correspondence for conformal field theories with boundaries (AdS/BCFT
correspondence) \cite{Taka11,Fuji11} and the geometric procedure for the
evaluation of the entanglement entropy for a bounded region in CFT \cite%
{Ryu06,Ryu06b}. In this section, based on \cite{Beze15}, we will consider
the effects on the scalar vacuum induced by a brane perpendicular to the AdS
boundary.

The background geometry is described by the line element (\ref{metric2}) and
the brane is located at $x^{1}=0$. The problem is symmetric with respect to
the brane and we will consider the region $x^{1}\geq 0$. The scalar field $%
\varphi (x)$ obeys the field equation (\ref{fieldeq}) and the boundary
condition
\begin{equation}
\left( \beta \partial _{1}+1\right) \varphi (x)=0,  \label{Rob2}
\end{equation}%
on the brane. Introducing the notations $\mathbf{x}=(x^{2},\ldots ,x^{D-1})$
and $\mathbf{k}=(k_{2},\ldots ,k_{D-1})$, the normalized mode functions
obeying the boundary condition have the form
\begin{equation}
\varphi _{\sigma }^{(\pm )}(x)=\frac{\sqrt{2k_{1}/\omega }}{\left( 2\pi
a\right) ^{(D-1)/2}}z^{D/2}J_{\nu }(\lambda z)\cos [k_{1}x^{1}+\alpha
_{0}(k_{1})]e^{i\mathbf{kx}\mp i\omega t},  \label{phisig2}
\end{equation}%
where $0\leq k_{1},\lambda <\infty $, the energy is given by $\omega =\sqrt{%
k^{2}+k_{1}^{2}+\lambda ^{2}}$ with $k^{2}=\sum_{i=2}^{D}k_{i}^{2}$ and $\nu
$ is defined by (\ref{nu}). The function $\alpha _{0}(k_{1})$ is determined
from the relation%
\begin{equation}
e^{2i\alpha _{0}(k_{1})}=\frac{i\beta k_{1}-1}{i\beta k_{1}+1}.  \label{alf0}
\end{equation}%
Note that in the case $0\leq \nu <1$ we impose the same boundary condition
on the AdS boundary as that in section \ref{sec:Modes} for the L-region.

For positive values of the Robin parameter $\beta $ in (\ref{Rob2}), in
addition to (\ref{phisig2}), there is a mode for which the part depending on
the coordinate $x^{1}$ is expressed in terms of the exponential function $%
e^{-x^{1}/\beta }$. The corresponding energy is given by $\omega =\sqrt{%
k^{2}+\lambda ^{2}-1/\beta ^{2}}$. In the subspace $k^{2}+\lambda
^{2}<1/\beta ^{2}$ of the quantum numbers the energy is imaginary and the
vacuum state is unstable. Note that for a\ Robin plate in the Minkowski bulk
the energy for the corresponding bound states is positive in the range $%
\beta >1/m$. In order to have a stable vacuum in the AdS bulk we will assume
that $\beta \leq 0$.

With the mode functions (\ref{phisig2}), the VEV of the energy-momentum
tensor is evaluated by using the mode-sum formula (\ref{Tmusum}) (for the
procedure based on the point-splitting regularization technique see \cite%
{Beze15}). The mode-sum contains the integration over the region $k_{1}\in
\lbrack 0,\infty )$. In the integrand, we write the parts containing the
products of the trigonometric functions with the arguments $%
k_{1}x^{1}+\alpha _{0}(k_{1})$ in the form of the sum of three terms. The
first one does not depend on $x^{1}$ and its contribution corresponds to the
VEV in the AdS spacetime when the brane is absent. The second and third
terms depend on the $x^{1}$-coordinate in the form of the exponents $%
e^{2ik_{1}x^{1}}$ and $e^{-2ik_{1}x^{1}}$. By taking into account that those
terms exponentially decay in the upper and lower half-planes of the complex
variable $k_{1}$, we rotate the integration contour over $k_{1}\in \lbrack
0,\infty )$ by the angle $\pi /2$ for the part with the exponent $%
e^{2ik_{1}x^{1}}$ and by the angle $-\pi /2$ for the part with $%
e^{-2ik_{1}x^{1}}$. The VEV of the energy-momentum tensor is presented in
the decomposed form (\ref{Tmudec}), where the diagonal components of the
brane-induced part are given by the expression (no summation over $\mu $)%
\begin{eqnarray}
\langle T_{\mu }^{\mu }\rangle _{\mathrm{b}}^{\mathrm{(s)}} &=&-\frac{\left(
4\pi \right) ^{(1-D)/2}}{2^{2\nu +1}a^{D+1}}\int_{0}^{\infty
}dx\,xe^{-2xx^{1}/z}\frac{\beta x/z+1}{\beta x/z-1}  \notag \\
&&\times \left[ A_{\mu }x^{D+2\nu }F_{\nu }^{D/2+1}(x)+\hat{B}_{\mu
}(x)x^{D+2\nu }F_{\nu }^{D/2}(x)\right] ,  \label{Tmus2}
\end{eqnarray}%
with $A_{1}=0$, $A_{D}=(1-D)/2$, $A_{l}=1/2$ for $l=0,2,\ldots ,D-1$. The
operators $\hat{B}_{\mu }(x)$ are defined as
\begin{eqnarray}
\hat{B}_{l}(x) &=&\frac{\xi _{1}}{4}\hat{B}(x)+\frac{\xi }{x}\left( \partial
_{x}-\frac{D}{x}\right) -\xi _{1}\delta _{1l},\;l=0,1,2,\ldots ,D-1,  \notag
\\
\hat{B}_{D}(x) &=&\frac{1}{4}\hat{B}(x)-D\frac{\xi }{x}\left( \partial _{x}-%
\frac{D}{x}\right) -\frac{m^{2}a^{2}}{x^{2}}+\xi _{1},  \label{BD}
\end{eqnarray}%
with $\hat{B}(x)=\partial _{x}^{2}-((D-1)/x)\partial _{x}+4$ and $\xi _{1}$
is given by (\ref{ksi1}). In (\ref{Tmus2}) we have introduced the function%
\begin{eqnarray}
F_{\nu }^{\mu }(x) &=&\frac{2^{2\nu +1}x^{-2\nu }}{\Gamma (\mu -1/2)}%
\int_{0}^{1}du\,u(1-u^{2})^{\mu -3/2}J_{\nu }^{2}(xu)  \notag \\
&=&\frac{\,_{1}F_{2}\left( \nu +1/2;\nu +\mu +1/2,2\nu +1;-x^{2}\right) }{%
\Gamma (\nu +\mu +1/2)\Gamma (\nu +1)},  \label{Fmunu2}
\end{eqnarray}%
where $_{1}F_{2}(a;b,c;y)$ is the hypergeometric function. The diagonal
components in the region $x^{1}<0$ are given by the expression (\ref{Tmus2})
with $x^{1}$ replaced by $|x^{1}|$.

An important difference from the geometry with a brane parallel to the AdS
boundary is the presence of the off-diagonal component of the vacuum
energy-momentum tensor:%
\begin{equation}
\langle T_{D}^{1}\rangle _{\mathrm{b}}^{\mathrm{(s)}}=-\frac{\left( 4\pi
\right) ^{(1-D)/2}}{2^{2\nu +2}a^{D+1}}\int_{0}^{\infty }dx\,e^{-2xx^{1}/z}%
\frac{\beta x/z+1}{\beta x/z-1}\left( \xi _{1}x\partial _{x}+4\xi \right)
x^{D+2\nu }F_{\nu }^{D/2}(x).  \label{T1D}
\end{equation}%
This expression for the off-diagonal component in the region $x^{1}<0$ is
obtained from (\ref{T1D}) changing the sign and replacing $x^{1}\rightarrow
|x^{1}|$. Note that $\langle T_{D}^{1}\rangle _{\mathrm{b}}^{\mathrm{(s)}%
}=\langle T_{1}^{D}\rangle _{\mathrm{b}}^{\mathrm{(s)}}$. Due to the nonzero
off-diagonal component $\langle T_{D}^{1}\rangle _{\mathrm{b}}^{\mathrm{(s)}%
} $, the Casimir force acting on the brane has two components. The first one
is determined by the stress $\langle T_{1}^{1}\rangle _{\mathrm{b}}^{\mathrm{%
(s)}}$ and corresponds to the component normal to the brane. The second one
is obtained from $\langle T_{D}^{1}\rangle _{\mathrm{b}}^{\mathrm{(s)}}$ and
is directed along the $z$-axis. It corresponds to the shear force. Of
course, because of the surface divergences in the local VEVs, both these
components require an additional renormalization. Note that the Casimir
forces acting tangential to the boundaries (lateral Casimir forces) may
arise also in condensed matter systems if the properties of the
corresponding surfaces are anisotropic or inhomogeneous (see, for example,
\cite{Chiu09,Chiu10} and references therein). In the problem under
consideration the tangential force is a consequence of the $z$-dependence of
the background geometry.

From the expressions (\ref{Tmus2}) and (\ref{T1D}) it follows that for the
Dirichlet and Neumann boundary conditions the brane-induced contributions to
the VEV of the energy-momentum tensor differ by the signs. In these special
cases the corresponding expressions are further simplified as (no summation
over $\mu $)%
\begin{eqnarray}
\langle T_{\mu }^{\mu }\rangle _{\mathrm{b}}^{\mathrm{(s)}} &=&\pm \frac{%
a^{-D-1}}{2^{D/2+\nu +1}\pi ^{D/2}}[\hat{C}_{\mu }(u)-D\xi ]f_{\nu }(u),
\notag \\
\langle T_{D}^{1}\rangle _{\mathrm{b}}^{\mathrm{(s)}} &=&\pm \frac{%
2a^{-D-1}x^{1}/z}{2^{D/2+\nu +1}\pi ^{D/2}}\left[ \xi _{1}\left( u-1\right)
\partial _{u}^{2}+\left( 2\xi -1\right) \partial _{u}\right] f_{\nu }(u),
\label{T1DN}
\end{eqnarray}%
where the upper and lower signs correspond to the Dirichlet and Neumann
boundary conditions, respectively, and $u=1+2(x^{1}/z)^{2}$. The operators $%
\hat{C}_{\mu }(u)$ in the expressions for the diagonal components are given
by%
\begin{equation}
\hat{C}_{l}(u)=\xi _{1}\left( u^{2}-1\right) \partial _{u}^{2}+\left[ 4\xi
-2+\left( \frac{D+1}{2}\xi _{1}-\frac{1}{2}\right) \left( u-1\right) \right]
\partial _{u},  \label{Cl}
\end{equation}%
for $l=0,2,\ldots ,D-1$, and
\begin{eqnarray}
\hat{C}_{1}(u) &=&\xi _{1}\left( u-1\right) ^{2}\partial _{u}^{2}+\left(
\frac{D+1}{2}\xi _{1}-\frac{1}{2}\right) \left( u-1\right) \partial _{u},
\notag \\
\hat{C}_{D}(u) &=&2\xi _{1}\left( u-1\right) \partial _{u}^{2}+\left[ 4\xi
-2+\frac{D}{2}\xi _{1}\left( u-1\right) \right] \partial _{u}.  \label{CD}
\end{eqnarray}%
The function $f_{\nu }(u)$ is expressed in terms of the hypergeometric
function $\,{}_{2}F_{1}\left( a,b;c;x\right) $:
\begin{equation}
f_{\nu }(u)=\frac{\Gamma (\nu +D/2)}{\Gamma (\nu +1)u^{\nu +D/2}}%
\,{}_{2}F_{1}\left( \frac{D+2\nu +2}{4},\frac{D+2\nu }{4};\nu +1;\frac{1}{%
u^{2}}\right) .  \label{fnu}
\end{equation}

For a conformally coupled massless field one has $\nu =1/2$ and the problem
under consideration is conformally related to the problem in the Minkowski
spacetime with the line element $ds^{2}=\eta _{\mu \rho }dx^{\mu }dx^{\rho }$%
, $x^{D}=z$, and with planar codimension one boundaries located at $z=0$
(the conformal image of the AdS boundary) and $x^{1}=0$ (the conformal image
of the brane). The Minkowskian field obeys the Dirichlet boundary condition
at $z=0$ and the Robin boundary condition (\ref{Rob2}) at $x^{1}=0$. The
Dirichlet boundary condition at $z=0$ is a consequence of the condition we
have imposed on the AdS boundary. Taking $\nu =1/2$, from the results given
above we can obtain the VEV of the energy-momentum tensor for a conformally
coupled massless field in the geometry of perpendicular planar boundaries in
the Minkowski bulk by using the relation $\left\langle T_{\mu }^{\rho
}\right\rangle _{\mathrm{b}}^{\mathrm{(M)}}=(a/z)^{D+1}\left\langle T_{\mu
}^{\rho }\right\rangle _{\mathrm{b}}$ (see (\ref{ConfRel})). In the
Minkowskian limit we get the result \ref{TmuM} (with the replacement $%
y-y_{0}\rightarrow x^{1}$) for the components $\mu =0,2,\ldots ,D$.

Near the brane and for non-Dirichlet boundary conditions, $x^{1}\ll z,|\beta
|$, the leading term in the asymptotic expansions for diagonal components
with $\mu =0,2,\ldots ,D$ is given by (no summation over $\mu $)
\begin{equation}
\langle T_{\mu }^{\mu }\rangle _{\mathrm{b}}^{\mathrm{(s)}}\approx -\frac{%
D\Gamma ((D+1)/2)\left( \xi -\xi _{D}\right) }{2^{D}\pi ^{(D+1)/2}\left(
ax^{1}/z\right) ^{D+1}}.  \label{Tmunear2}
\end{equation}%
This coincides with the Minkowskian result where the distance from the
boundary is replaced by the ratio $ax^{1}/z$ (compare with (\ref{T00nearS}%
)). Note that, in accordance with (\ref{metric2}), the latter is the proper
distance from the brane measured by an observer at rest with respect to the
brane. Note that the proper distance $ax^{1}/z$ is different from the
geodesic distance $\sigma (x,x^{\prime })$. The latter between the spacetime
points $x=(t,0,\mathbf{x},z)$ and $x^{\prime }=(t,x^{1},\mathbf{x},z)$ is
given as $\cosh (\sigma (x,x^{\prime })/a)=1+(x^{1}/z)^{2}/2$. For the
normal stress and for the off-diagonal component near the brane one gets%
\begin{equation}
\langle T_{1}^{1}\rangle _{\mathrm{b}}^{\mathrm{(s)}}\approx -\frac{\left(
x^{1}/z\right) ^{2}}{D-1}\langle T_{0}^{0}\rangle _{\mathrm{b}}^{\mathrm{(s)}%
},\;\langle T_{D}^{1}\rangle _{\mathrm{b}}^{\mathrm{(s)}}\approx \frac{x^{1}%
}{z}\langle T_{0}^{0}\rangle _{\mathrm{b}}^{\mathrm{(s)}}.  \label{T11near}
\end{equation}%
The leading terms for the Dirichlet boundary condition differ from the ones
given above by the sign. The expressions (\ref{Tmunear2}) and (\ref{T11near}%
) describe also the asymptotic behavior of the brane-induced VEV near the
AdS horizon (large values of $z$ for fixed $x^{1}$).

Now let us consider the asymptotics at large distances from the brane, $%
x^{1}\gg z$. For non-Neumann boundary conditions, additionally assuming that
$x^{1}\gg |\beta |$, for the components $\mu =0,\ldots ,D-1$ the leading
term is given by (no summation over $\mu $)
\begin{equation}
\langle T_{\mu }^{\mu }\rangle _{\mathrm{b}}^{\mathrm{(s)}}\approx \frac{%
\left( \xi -1/4\right) \left( D+2\nu \right) +\xi }{\pi ^{D/2}\Gamma (\nu
)a^{D+1}(2x^{1}/z)^{D+2\nu }}\Gamma (D/2+\nu ).  \label{TmuLarge0}
\end{equation}%
The asymptotics for the remaining components are expressed as%
\begin{equation}
\langle T_{D}^{D}\rangle _{\mathrm{b}}^{\mathrm{(s)}}\approx -\frac{D}{2\nu }%
\langle T_{0}^{0}\rangle _{\mathrm{b}}^{\mathrm{(s)}},\;\langle
T_{D}^{1}\rangle _{\mathrm{b}}^{\mathrm{(s)}}\approx \frac{D/2+\nu }{\nu
x^{1}/z}\langle T_{0}^{0}\rangle _{\mathrm{b}}^{\mathrm{(s)}}.
\label{TmuLarge}
\end{equation}%
As seen, the leading order terms for $0<|\beta |<\infty $ coincide with
those for the Dirichlet boundary condition. For a scalar field with the
Neumann boundary condition ($\beta =0$), the leading terms differ from those
for the Dirichlet condition by the signs. Hence, the Dirichlet boundary
condition is the attractor in a class of Robin boundary conditions with $%
\beta \neq 0$. At large distances, the brane-induced contributions,
considered as functions of the proper distance from the brane, exhibit a
power-law fall-off for both massless and massive fields. This behavior is in
clear contrast with the case of the Minkowski bulk, where the
boundary-induced VEVs decay exponentially for massive fields, like $%
e^{-2mx^{1}}$ (see (\ref{TmuM})). Note that for large $x^{1}/z$ one has the
relation $(x^{1}/z)^{2}=\exp [\sigma (x,x^{\prime })/a]$, with $\sigma
(x,x^{\prime })$ being the geodesic distance. For fixed $x^{1}$, the
asymptotic formulas (\ref{TmuLarge0}) and (\ref{TmuLarge}) describe the
behavior of the VEVs near the AdS boundary. The diagonal components decay as
$z^{D+2\nu }$, whereas the off-diagonal component tends to zero like $%
z^{D+2\nu +1}$. The qualitative behavior of the brane-induced energy
density, as a function of the distance from the brane, is similar to that we
have described in the previous section for a scalar field with the Robin
boundary condition on the brane parallel to the AdS boundary.

\section{Summary}

\label{sec:Conc}

We have considered the influence of a brane in AdS bulk on the properties of
quantum vacuum. Two geometries are discussed: (i) a brane parallel to the
AdS boundary and (ii) a brane perpendicular to the AdS boundary. In the
first geometry, as a local characteristic of the vacuum state, the VEV of
the energy-momentum tensor is investigated for scalar, Dirac and
electromagnetic fields. For scalar field a general Robin boundary condition
is considered and the Dirac field is constrained by the bag boundary
condition. In the case of the electromagnetic field two types of boundary
conditions are discussed. The first one corresponds to the perfect conductor
boundary conditions in 3D electrodynamics and the second one is the analog
of the boundary condition used in bag models of hadrons to confine the
gluons. The VEV of the energy-momentum tensor is expressed as a mode-sum
over complete set of mode functions and for all these cases the
corresponding sets are given. The brane divides the background geometry into
two regions: the region between the brane and AdS horizon (R-region) and the
region between the brane and AdS boundary (L-region). Though the AdS
spacetime is homogeneous, the brane has a nonzero extrinsic curvature tensor
and the properties of the quantum vacuum in those regions are different. In
particular, the spectrum of the quantum number $\lambda $, corresponding to
the momentum along the direction normal to the AdS boundary, is continuous
in the R-region and discrete in the L-region. In the latter region the
eigenvalues are zeros of cylinder functions. The mode-sum for the VEV of the
energy-momentum tensor contains series over those zeros and for the
summation we have employed the generalized Abel-Plana formula. That allowed
to extract from the VEV the part corresponding to the geometry without a
brane and to present the brane-induced contribution in terms of integral,
exponentially convergent for points away form the brane. A similar
decomposition is provided for the R-region.

Near the brane, the leading terms in the asymptotic expansions for the
energy density and parallel stresses coincide with the corresponding
expressions for a single boundary in the Minkowski bulk, where the distance
from the boundary is replaced by the proper distance from the brane on the
AdS bulk. For those VEVs the effect of gravity is weak. This is related to
the fact that near the brane the main contribution to the corresponding VEVs
come from the vacuum fluctuations with the wavelengths smaller than the
curvature radius of the background geometry and influence of the
gravitational field on those modes is weak. For a boundary in the Minkowski
bulk the normal stress is zero. The nonzero normal stress in the geometry of
a brane on the AdS bulk is a purely gravitational effect. The effect of
gravity on the brane-induced VEVs is essential at distances from the brane
larger than the curvature radius. In particular, for the R-region, at large
distances the decay of the brane-induced contribution in the vacuum
energy-momentum tensor, as a function of the proper distance, is exponential
for both massless and massive fields. For the Minkowski bulk and for
massless fields the fall-off of the boundary-induced contribution is as
power-law. On the AdS boundary, the brane-induced contributions tend to zero
like $z^{D+\beta }$, where $\beta =2\nu $ for scalar field (with $\nu $
given by (\ref{nu})), $\beta =2ma+1$ for the Dirac field, and $\beta =D-2$
for the electromagnetic field. Near the AdS boundary one has a simple
relation between the energy density and the normal stress, given by $\langle
T_{D}^{D}\rangle _{\mathrm{b}}\approx -(D/\beta )\langle T_{0}^{0}\rangle _{%
\mathrm{b}}$. This correspond to the barotropic equation of state for the
vacuum pressure $-\langle T_{D}^{D}\rangle _{\mathrm{b}}$ along the $z$%
-direction and vacuum energy density. Note that for the pressures along the
directions parallel to the brane the equation of state is of the
cosmological constant type. By using the generalized zeta function
technique, we have also investigated the VEV\ of the surface energy-momentum
tensor. From the viewpoint of the observer living on the brane, the latter
corresponds to a gravitational source of cosmological constant type.
Depending on the value of the coefficient in the boundary condition, the
induced cosmological constant can be either positive and negative.

The brane-induced effects on the quantum vacuum for the geometry (ii) we
have considered on the example of a scalar field with general curvature
coupling parameter. For the Robin boundary condition the mode functions have
the form (\ref{phisig2}). The diagonal components of the brane-induced
energy-momentum tensor are given by the expressions (\ref{Tmus2}). An
important difference from the problem with a brane parallel to the AdS
boundary is the presence of nonzero off-diagonal component (\ref{T1D}) of
the vacuum energy-momentum tensor. As a consequence, the Casimir force
acting on the brane, in addition to the normal component, contains a
component directed parallel to the brane (shear force). At large distances
from the brane, the decay of the brane-induced contribution to the
energy-momentum tensor, as a function of the proper distance from the brane,
is as power-law for both massive and massless field. As it has been
mentioned above, in the Minkowski bulk the decay for massive fields is
exponential.

For charged fields, another important local characteristic of the vacuum
state is the expectation value of the current density. This VEV in models
with local AdS geometry and with toroidally compact spatial dimensions, in
the presence of single and two branes has been investigated in \cite%
{Bell15,Bell16Cb} and \cite{Bell18,Bell20} for charged scalar and fermionic
fields. The vacuum currents have nonzero components along the compact
dimensions only. They are periodic functions of the magnetic flux with the
period equal to the flux quantum. Depending on the boundary conditions
imposed on the fields at the locations of the branes, the brane-induced
effects lead to increase or decrease of the current density. Applications
were discussed to Randall-Sundrum type braneworld models and also to curved
graphene tubes.

In the discussion above we have assumed that the background geometry is
fixed. Among the interesting directions for the further research is the
investigation of the back-reaction of quantum effects on the geometry by
using the semiclassical Einstein equations with the VEV of the
energy-momentum tensor in the right-hand side. The vacuum energy-momentum
tensor may violate the energy conditions in the singularity theorems and
this leads to interesting comsological dynamics of the bulk and on the
brane. In this regard, the next step in the study of local quantum effects
in braneworlds could be the investigation of the vacuum energy-momentum
tensor in models with dS branes.

\section*{Acknowledgments}

The author is grateful to Stefano Bellucci, Eugenio Bezerra de Mello, Emilio
Elizalde, Sergei Odintsov, Mohammad Setare and Valery Vardanyan for
collaboration.


\begin{thebibliography}{999}
\bibitem{Call90} Callan, C.G.; Wilczek, F. Infrared behavior at negative
curvature. \textit{Nucl. Phys. B }\textbf{1990}\textit{,} \textit{340},
366-386.

\bibitem{Ahar00} Aharony, O.; Gubser, S.S.; Maldacena, J.; Ooguri, H.; Oz,
Y. Large N field theories, string theory and gravity. \textit{Phys. Rep. }%
\textbf{2000,}\textit{\ 323}, 183-386.

\bibitem{Nast15} N\u{a}stase, H. \textit{Introduction to AdS/CFT
correspondence;} Cambridge University Press: Cambridge, UK, 2015.

\bibitem{Ammo15} Ammon, M.; Erdmenger, J. \textit{Gauge/Gravity Duality:
Foundations and Applications;} Cambridge University Press: Cambridge, UK,
2015.

\bibitem{Pire14} Pires, A.S.T. \textit{AdS/CFT Correspondence in Condensed
Matter}; Morgan \& Claypool Publishers: San Rafael, CA, USA, 2014.

\bibitem{Zaan15} Zaanen, J.; Sun, Y.-W.; Liu, Y.; Schalm, K. \textit{%
Holographic Duality in Condensed Matter Physics;} Cambridge University
Press: Cambridge, UK, 2015.

\bibitem{Maar10} Maartens, R.; Koyama, K. Brane-World Gravity. \textit{%
Living Rev. Relativity, }\textbf{2010}\textit{, 13}, 1.

\bibitem{Most97} Mostepanenko, V.M.; Trunov, N.N. \textit{The Casimir Effect
and Its Applications}; Clarendon Press: Oxford, UK, 1997.

\bibitem{Milt02} Milton, K.A. \textit{The Casimir Effect: Physical
Manifestation of Zero-Point Energy;} World Scientific: Singapore, 2002.

\bibitem{Bord09} Bordag, M.; Klimchitskaya, G.L.; Mohideen, U.;
Mostepanenko, V.M. \textit{Advances in the Casimir Effect;} Oxford
University Press: New York, USA, 2009.

\bibitem{Casi11} Dalvit, D.; Milonni, P.; Roberts, D.; da Rosa, F. (Eds.)
\textit{Casimir Physics}; Lecture Notes in Physics; Springer-Verlag: Berlin,
Germany, 2011; Volume 834.

\bibitem{Gold99} Goldberger, W.D.; Wise, M.B. Modulus stabilization with
bulk fields. \textit{Phys. Rev. Lett.} \textbf{1999}, \textit{83}, 4922-4925.

\bibitem{Gold00a} Goldberger, W.D.; Wise, M.B. Phenomenology of a stabilized
modulus. \textit{Phys. Lett. B} \textbf{2000}, \textit{475}, 275-279.

\bibitem{Lesg04} Lesgourgues, J.; Sorbo, L. Goldberger-Wise variations:
stabilizing brane models with a bulk scalar. \textit{Phys. Rev. D} \textbf{%
2004}, \textit{69}, 084010.

\bibitem{Chou13} Choudhury, S.; SenGupta, S. Features of warped geometry in
presence of Gauss-Bonnet coupling. \textit{J. High Energy Phys.} \textbf{2013%
}, \textit{02}, 136.

\bibitem{Flac13} Flachi, A; Minamitsuji, M.; Uzawa, K. Moduli stabilization
in a de Sitter compactification model. \textit{J. High Energy Phys.} \textbf{%
2013}, \textit{08}, 073.

\bibitem{Chou14} Choudhury, S,; Mitra, J.; SenGupta, S. Modulus
stabilization in higher curvature dilaton gravity. \textit{J. High Energy
Phys.} \textbf{2014}, \textit{08}, 004.

\bibitem{Fuji20} Fujikura, K.; Nakai, Y.; Yamada, M. A more attractive
scheme for radion stabilization and supercooled phase transition. \textit{J.
High Energy Phys.} \textbf{2020}, \textit{02}, 111.

\bibitem{Fabi00} Fabinger, M.; Horava, P. Casimir effect between
world-branes in heterotic M-theory. \textit{Nucl. Phys. B} \textbf{2000},
\textit{580}, 243-263.

\bibitem{Noji00} Nojiri, S.; Odintsov, S.; Zerbini, S. Quantum (in)stability
of dilatonic AdS backgrounds and the holographic renormalization group with
gravity. \textit{Phys. Rev. D} \textbf{2000}, \textit{62}, 064006.

\bibitem{Toms00} Toms, D.J. Quantised bulk fields in the Randall-Sundrum
compactification model. \textit{Phys. Lett. B} \textbf{2000}, \textit{484},
149-153.

\bibitem{Noji00b} Nojiri, S.; Obregon, O.; Odintsov, S. (Non)-singular
brane-world cosmology induced by quantum effects in five-dimensional
dilatonic gravity. \textit{Phys. Rev. D} \textbf{2000}, \textit{62}, 104003.

\bibitem{Gold00} Goldberger, W.D.; Rothstein, I.Z. Quantum stabilization of
compactified AdS$_{5}$. \textit{Phys. Lett. B} \textbf{2000}, \textit{491},
339-344.

\bibitem{Noji00c} Nojiri, S.; Odintsov, S. Brane-world cosmology in higher
derivative gravity or warped compactification in the next-to-leading order
of AdS/CFT correspondence. \textit{J. High Energy Phys.} \textbf{2000},%
\textit{07}, 049.

\bibitem{Garr01} Garriga, J.; Pujol\`{a}s, O.; Tanaka, T. Radion effective
potential in the brane-world. \textit{Nucl. Phys. B} \textbf{2001}, \textit{%
605}, 192-214.

\bibitem{Flac01} Flachi, A.; Toms, D.J. Quantized bulk scalar fields in the
Randall-Sundrum brane model. \textit{Nucl. Phys. B} \textbf{2001}, \textit{%
610}, 144-168.

\bibitem{Brev01} Brevik, I.H.; Milton, K.A.; Nojiri, S.; Odintsov, S.D.
Quantum (in)stability of a brane-world $AdS_{5}$ universe at nonzero
temperature. \textit{Nucl. Phys. B} \textbf{2001}, \textit{599}, 305-318.

\bibitem{Saha03} Saharian, A.A.; Setare, M.R. The Casimir effect on
background of conformally flat brane-world geometries. \textit{Phys. Lett. B}
\textbf{2003}, \textit{552}, 119-126.

\bibitem{Yera03} Yeranyan, A.H.; Saharian, A.A. Cosmological dynamics of
brane models and vacuum effects. \textit{Astrophysics} \textbf{2003},
\textit{46}, 386-397.

\bibitem{Durr07} Durrer, R.; Ruser, M. Dynamical Casimir Effect in
Braneworlds. \textit{Phys. Rev. Lett.} \textbf{2007}, \textit{99}, 071601.

\bibitem{Ruse07} Ruser, M.; Durrer, R. Dynamical Casimir effect for
gravitons in bouncing braneworlds. \textit{Phys. Rev. D} \textbf{2007},
\textit{76}, 104014.

\bibitem{Frank07} Frank, M.; Turan, I., Ziegler, L. Casimir force in
Randall-Sundrum models. \textit{Phys. Rev. D} \textbf{2007}, \textit{76},
015008.

\bibitem{Flac09} Flachi, A.; Tanaka, T. Casimir effect on the brane. \textit{%
Phys. Rev. D} \textbf{2009}, \textit{80}, 124022.

\bibitem{Teo09} Teo, L.P. Casimir effect in spacetime with extra dimensions
- from Kaluza-Klein to Randall-Sundrum models. \textit{Phys. Lett. B}
\textbf{2009}, \textit{682}, 259-265.

\bibitem{Rype10} Rypestol, M.; Brevik, I. Finite-temperature Casimir effect
in Randall-Sundrum models. \textit{New. J. Phys.} \textbf{2010}, \textit{12}%
, 013022.

\bibitem{Obou11} Obousy, R.; Cleaver, G. Casimir energy and brane stability.
\textit{J. Geom. Phys.} \textbf{2011}, \textit{61}, 577-588.

\bibitem{Teo13} Teo, L.P. Finite temperature fermionic Casimir interaction
in Anti-de Sitter Space-time. \textit{Int. J. Mod. Phys. A} \textbf{2013},
\textit{28}, 1350158.

\bibitem{Haba19} Haba, N.; Yamada, T. Revisiting quantum stabilization of
the radion in Randall-Sundrum model. arXiv: 1903.10160.

\bibitem{Flac01b} Flachi, A.; Moss, I.G.; Toms, D.J. Fermion vacuum energies
in brane world models. \textit{Phys. Lett. B} \textbf{2001}, \textit{518},
153-156.

\bibitem{Flac01c} Flachi, A.; Moss, I.G.; Toms, D.J. Quantized bulk fermions
in the Randall-Sundrum brane model. \textit{Phys. Rev. D} \textbf{2001},
\textit{64}, 105029.

\bibitem{Uzav03} Uzawa, K. Dilaton Stabilization in (A)dS Spacetime with
Compactified Dimensions. \textit{Prog. Theor. Phys.} \textbf{2003}, \textit{%
110}, 457-498.

\bibitem{Shao10} Shao, S.-H.; Chen, P.; Gu, J.-A. Stress-energy tensor
induced by a bulk Dirac spinor in the Randall-Sundrum model. \textit{Phys.
Rev. D} \textbf{2010}, \textit{81}, 084036.

\bibitem{Eliz13} Elizalde, E.; Odintsov, S.D.; Saharian, A.A. Fermionic
Casimir densities in anti-de Sitter spacetime. \textit{Phys. Rev. D} \textbf{%
2013}, \textit{87}, 084003.

\bibitem{Garr03} Garriga, J.; Pomarol, A. A stable hierarchy from Casimir
forces and the holographic interpretation. \textit{Phys. Lett. B} \textbf{%
2003}, \textit{560}, 91-97.

\bibitem{Teo10} Teo, L.P. Casimir effect of electromagnetic field in
Randall-Sundrum spacetime. \textit{J. High Energy Phys.} \textbf{2010},
\textit{10}, 019.

\bibitem{Saha16} Saharian, A.A.; Kotanjyan, A.S.; Saharyan, A.A. \textit{%
Proceedings of the Yerevan State University. Phys. Math.} \textbf{2016},
\textit{3}, 37-41.

\bibitem{Kota17} Kotanjyan, A.S.; Saharian, A.A. Electromagnetic quantum
effects in anti-de Sitter spacetime. \textit{Physics of Atomic Nuclei}
\textbf{2017}, \textit{80}, 562-571.

\bibitem{Kota17b} Kotanjyan, A.S.; Saharian, A.A.; Saharyan, A.A.
Electromagnetic Casimir Effect in AdS Spacetime. \textit{Galaxies} \textbf{%
2017}, \textit{5}, 102.

\bibitem{Saha20} Saharian, A.A.; Kotanjyan, A.S.; Sargsyan, H.G.
Electromagnetic field correlators and the Casimir effect for planar
boundaries in AdS spacetime with application in braneworlds;
arXiv:2009.02072.

\bibitem{Noji00d} Nojiri, S.; Odintsov, S. Brane world inflation induced by
quantum effects. \textit{Phys. Lett. B} \textbf{2000}, \textit{484}, 119-123.

\bibitem{Nayl02} Naylor, W.; Sasaki, M. Casimir energy for de Sitter branes
in bulk $AdS_{5}$. \textit{Phys. Lett. B} \textbf{2002}, \textit{542},
289-294.

\bibitem{Eliz03} Elizalde, E.; Nojiri, S.; Odintsov, S.D.; Ogushi, S.
Casimir effect in de Sitter and anti-de Sitter braneworlds. \textit{Phys.
Rev. D} \textbf{2003}, \textit{67}, 063515.

\bibitem{Moss03} Moss, I.G.; Naylor, W.; Santiago-Germ\'{a}n, W.; Sasaki, M.
Bulk quantum effects for de Sitter branes in $AdS_{5}$. \textit{Phys. Rev. D}
\textbf{2003}, \textit{67}, 125010.

\bibitem{Pujo04} Pujol\`{a}s, O.; Tanaka, T. Massless scalar fields and
infrared divergences in the inflationary brane world. \textit{J. Cosmol.
Astropart. Phys.} \textbf{2004}, \textit{12}, 009.

\bibitem{Flac04} Flachi, A.; Knapman, A.; Naylor, W.; Sasaki, M. Zeta
functions in brane world cosmology. \textit{Phys. Rev. D} \textbf{2004},
\textit{70}, 124011.

\bibitem{Norm04} Norman, J.P. Casimir effect between anti-de Sitter
braneworlds. \textit{Phys. Rev. D} \textbf{2004}, \textit{69}, 125015.

\bibitem{Nayl05} Naylor, W.; Sasaki, M. Quantum Fluctuations for de Sitter
Branes in Bulk $AdS_{5}$. \textit{Prog. Theor. Phys.} \textbf{2005}, \textit{%
113}, 535-554.

\bibitem{Pujo05} Pujol\`{a}s, O.; Sasaki, M. Vacuum destabilization from
Kaluza-Klein modes in an inflating brane. \textit{J. Cosmol. Astropart. Phys.%
} \textbf{2005}, \textit{09}, 002.

\bibitem{Flac03a} Flachi, A.; Garriga, J.; Pujol\`{a}s, O.; Tanaka, T.
Moduli stabilization in higher dimensional brane models. \textit{J. High
Energy Phys.} \textbf{2003}, \textit{08}, 053.

\bibitem{Flac03b} Flachi, A.; Pujol\`{a}s, O. Quantum self-consistency of AdS%
$\times \Sigma $ brane models. \textit{Phys. Rev. D} \textbf{2003},\textit{\
68}, 025023.

\bibitem{Saha06a} Saharian, A.A. Wightman function and vacuum fluctuations
in higher dimensional brane models. \textit{Phys. Rev. D} \textbf{2006,}
\textit{73}, 044012.

\bibitem{Saha06b} Saharian, A.A. Bulk Casimir densities and vacuum
interaction forces in higher dimensional brane models. \textit{Phys. Rev. D}
\textbf{2006,} \textit{73}, 064019.

\bibitem{Eliz07} Elizalde, E.; Minamitsuji, M.; Naylor, W. Casimir effect in
rugby-ball type flux compactifications. \textit{Phys. Rev. D} \textbf{2007},
\textit{75}, 064032.

\bibitem{Lina08} Linares, R.; Morales-T\'{e}cotl, H.A.; Pedraza, O. Casimir
force for a scalar field in warped brane worlds. \textit{Phys. Rev. D}
\textbf{2008}, \textit{77}, 066012.

\bibitem{Fran08} Frank, M.; Saad, N.; Turan, I. Casimir force in
Randall-Sundrum models with $q+1$ dimensions. \textit{Phys. Rev. D} \textbf{%
2008}, \textit{78}, 055014.

\bibitem{Knap04} Knapman, A.; Toms, D.J. Stress-energy tensor for a
quantized bulk scalar field in the Randall-Sundrum brane model. \textit{%
Phys. Rev. D }\textbf{2004,}\textit{\ 69}, 044023.

\bibitem{Saha05} Saharian, A.A. Wightman function and Casimir densities on
AdS bulk with application to the Randall-Sundrum braneworld. \textit{Nucl.
Phys. B} \textbf{2005}, \textit{712}, 196-228.

\bibitem{Saha07} Saharian, A.A.; Mkhitaryan, A.L. Wightman function and
vacuum densities for a Z$_{2}$-symmetric thick brane in AdS spacetime.
\textit{J. High Energy Phys.} \textbf{2007}, \textit{08}, 063.

\bibitem{Beze15} Bezerra de Mello, E.R.; Saharian, A.A.; Setare, M. R.
Vacuum densities for a brane intersecting the AdS boundary. \textit{Phys.
Rev. D} \textbf{2015,} \textit{92}, 104005.

\bibitem{Brei82} Breitenlohner, P.; Freedman, D.Z. Stability in gauged
extended supergravity. \textit{Ann. Phys. (NY)} \textbf{1982,} \textit{144
(2)}, 249-281.

\bibitem{Mezi85} Mezincescu, L.; Townsend, P.K. Stability at a local maximum
in higher dimensional anti-deSitter space and applications to supergravity.
\textit{Ann. Phys. (NY)} \textbf{1985}, \textit{160 (2)}, 406-419.

\bibitem{Bell15} Bellucci, S.; Saharian, A.A.; Vardanyan, V. Vacuum currents
in braneworlds on AdS bulk with compact dimensions. \textit{J. High Energy
Phys.} \textbf{2015}, \textit{11}, 092.

\bibitem{Avis78} Avis, S.J.; Isham, C.J.; Storey, D. Quantum field theory in
anti-de Sitter space-time. \textit{Phys. Rev. D} \textbf{1978,} \textit{18},
3565-3576.

\bibitem{Ishi04} Ishibashi, A.; Wald, R.M. Dynamics in
non-globally-hyperbolic static spacetimes: III. Anti-de Sitter spacetime.
\textit{Class. Quantum Grav.} \textbf{2004}, \textit{21}, 2981.

\bibitem{Morl20} Morley, T.; Taylor, P.; Winstanley, E. Quantum field theory
on global anti-de Sitter space-time with Robin boundary conditions;
arXiv:2004.02704.

\bibitem{Bell18} Bellucci, S.; Saharian, A.A.; Simonyan, D.H.; Vardanyan, V.
Fermionic currents in topologically nontrivial braneworlds. \textit{Phys.
Rev. D} \textbf{2018}, \textit{98}, 085020.

\bibitem{Rand99a} Randall, L.; Sundrum, R. Large mass hierarchy from a small
extra dimension. \textit{Phys. Rev. Lett.} \textbf{1999}, \textit{83},
3370-3373.

\bibitem{Rand99b} Randall, L.; Sundrum, R. An alternative to
compactification. \textit{Phys. Rev. Lett.} \textbf{1999}, \textit{83},
4690-4693.

\bibitem{Gher00} Gherghetta, T.; Pomarol, A. Bulk fields and supersymmetry
in a slice of AdS. \textit{Nucl. Phys. B} 2000, \textit{586}, 141-162.

\bibitem{Chan05} Chan, S.; Park, S.Ch.; Song, J. Kaluza-Klein masses of bulk
fields with general boundary conditions in AdS$_{5}$ space. \textit{Phys.
Rev. D} \textbf{2005}, \textit{71}, 106004.

\bibitem{Grib94} Grib, A.A.; Mamayev, S.G.; Mostepanenko, V.M. \textit{%
Vacuum Quantum Effects in Strong Fields;} Friedmann Laboratory Publishing:
St. Petersburg, 1994.

\bibitem{Birr82} Birrell, N.D.; Davies, P.C.W. \textit{Quantum Fields in
Curved Space;} Cambridge University Press: Cambridge, England, 1982.

\bibitem{Buch92} Buchbinder, I.L.; Odintsov, S.D.; Shapiro, I.L.\textit{\
Effective Action in Quantum Gravity;} Taylor \& Francis: BocaRaton, 1992.

\bibitem{Eliz94} Elizalde, E.; Odintsov, S.D.; Romeo, A.; Bytsenko, A.A.;
Zerbini, S. \textit{Zeta Regularization Techniques with Applications;} World
Scientific: Singapore, 1994.

\bibitem{Kirs01} Kirsten, K. \textit{Spectral Functions in Mathematics and
Physics;} CRC Press: Boca Raton, FL, 2001.

\bibitem{Byts03} Bytsenko, A.A.; Cognola, G.; Elizalde, E.; Moretti, V.;
Zerbini, S. \textit{Analytic Aspects of Quantum Fields;} World Scientific:
Singapore, 2003.

\bibitem{Kent15} Kent, C.; Winstanley, E. Hadamard renormalized scalar field
theory on anti-de Sitter spacetime. \textit{Phys. Rev. D} \textbf{2015},
\textit{91}, 044044.

\bibitem{Ambr15} Ambrus, V.E.; Winstanley, E. Renormalised fermion vacuum
expectation values on anti-deSitter space-time. \textit{Phys. Lett. B}
\textbf{2015}, \textit{749}, 597-602.

\bibitem{Prud86} Prudnikov, A.P.; Brychkov, Yu.A.; Marichev, O.I.~ \textit{%
Integrals and series;} Gordon and Breach: New York, 1986, Vol.2.

\bibitem{Saha87} Saharian, A.A. A generalized Abel-Plana formula.
Applications to cylindrical functions. \textit{Izv. Akad. Nauk Arm. SSR Mat.}
\textbf{1987}, \textit{22}, 166-179 [\textit{Sov. J. Contemp. Math. Anal.}
\textbf{1987}, \textit{22}, 70-86].

\bibitem{Saha08} Saharian, A.A. \textit{The Generalized Abel-Plana Formula
with Applications to Bessel Functions and Casimir Effect;} Yerevan State
University Publishing House: Yerevan, 2008; Report No. ICTP/2007/082;
arXiv:0708.1187.

\bibitem{Saha08dS} Saharian, A.A.; Vardanyan, T.A. Casimir densities for a
plate in de Sitter spacetime. \textit{Class. Quantum Grav.} \textbf{2009},
\textit{26}, 195004.

\bibitem{Eliz10dS} Elizalde, E.; Saharian, A.A.; Vardanyan, T.A. Casimir
effect for parallel plates in de Sitter spacetime. \textit{Phys. Rev. D}
\textbf{2010}, \textit{81}, 124003.

\bibitem{Saha14dS} Saharian, A.A.; Kotanjyan, A.S.; Nersisyan, H.A.
Electromagnetic two-point functions and Casimir densities for a conducting
plate in de Sitter spacetime. \textit{Phys. Lett. B} \textbf{2014}, \textit{%
728}, 141-147.

\bibitem{Saha15dS} Saharian, A.A.; Kotanjyan, A.S.; Nersisyan, H.A.
Electromagnetic Casimir effect for conducting plates in de Sitter spacetime.
\textit{Phys. Scr.} \textbf{2015}, \textit{90}, 065304.

\bibitem{Saha04Surf} Saharian, A.A. Energy-momentum tensor for a scalar
field on manifolds with boundaries. \textit{Phys. Rev. D} \textbf{2004},
\textit{69}, 085005.

\bibitem{Saha04b} Saharian, A.A. Surface Casimir densities and induced
cosmological constant on parallel branes in AdS spacetime \textit{Phys. Rev.
D} \textbf{2004}, \textit{70}, 064026.

\bibitem{Saha06} Saharian, A.A. Surface Casimir densities and induced
cosmological constant in higher dimensional braneworlds. \textit{Phys. Rev. D%
} \textbf{2006,} \textit{74}, 124009.

\bibitem{Saha18} Saharian, A.A.; Sargsyan, H. G. Induced Cosmological
Constant in Brane Models with a Compact Dimension. \textit{Astrophysics}
\textbf{2018}, \textit{61}, 375--390.

\bibitem{Taka11} Takayanagi, T. Holographic Dual of a Boundary Conformal
Field Theory. \textit{Phys. Rev. Lett.} \textbf{2011},\textit{\ 107}, 101602.

\bibitem{Fuji11} Fujita, M.; Takayanagi, T.; Tonni, E. Aspects of AdS/BCFT.
\textit{J. High Energy Phys.} \textbf{2011}, \textit{11}, 043.

\bibitem{Ryu06} Ryu, S; Takayanagi, T. Holographic Derivation of
Entanglement Entropy from the anti-de Sitter Space/Conformal Field Theory
Correspondence. \textit{Phys. Rev. Lett.} \textbf{2006}, \textit{96}, 181602.

\bibitem{Ryu06b} Ryu, S; Takayanagi, T. Aspects of holographic entanglement
entropy. \textit{J. High Energy Phys. }\textbf{2006}\textit{,} \textit{08},
045.

\bibitem{Chiu09} Chiu, H.-C.; Klimchitskaya, G.L.; Marachevsky, V.N.;
Mostepanenko, V.M.; Mohideen, U. Demonstration of the asymmetric lateral
Casimir force between corrugated surfaces in the nonadditive regime. \textit{%
Phys. Rev. B} \textbf{2009}, \textit{80}, 21402(R).

\bibitem{Chiu10} Chiu, H.-C.; Klimchitskaya, G.L.; Marachevsky, V.N.;
Mostepanenko, V.M.; Mohideen, U. Lateral Casimir force between sinusoidally
corrugated surfaces: Asymmetric profiles, deviations from the proximity
force approximation, and comparison with exact theory. \textit{Phys. Rev. B}
\textbf{2010}, \textit{81}, 115417.

\bibitem{Bell16Cb} Bellucci, S.; Saharian, A.A.; Vardanyan, V. Hadamard
function and the vacuum currents in braneworlds with compact dimensions:
Two-brane geometry. \textit{Phys. Rev. D} \textbf{2016}, \textit{93}, 084011.

\bibitem{Bell20} Bellucci, S.; Saharian, A.A.; Sargsyan, H.G.; Vardanyan,
V.V. Fermionic vacuum currents in topologically nontrivial braneworlds:
Two-brane geometry. \textit{Phys. Rev. D} \textbf{2020}, \textit{101},
045020.
\end{thebibliography}
\end{document}